# THE 2001 SUPEROUTBURST OF WZ SAGITTAE

JOSEPH PATTERSON,[1] GIANLUCA MASI,[2] MICHAEL W. RICHMOND,[3] BRIAN MARTIN,[4]
EDWARD BESHORE,[5] DAVID R. SKILLMAN,[6] JONATHAN KEMP,[1,7,8] TONNY VANMUNSTER,[9]
ROBERT REA,[10] WILLIAM ALLEN,[11] STACEY DAVIS,[3] TRACY DAVIS,[3] ARNE A. HENDEN,[12]
DONN STARKEY,[13] JERRY FOOTE,[14] ARTO OKSANEN,[15] LEWIS M. COOK,[16] ROBERT E. FRIED,[17]
DIETER HUSAR,[18] RUDOLF NOVÁK,[19] TUT CAMPBELL,[20] JEFF ROBERTSON,[20] THOMAS KRAJCI,[21]
ELENA PAVLENKO,[22] NESTOR MIRABAL,[1] PANOS G. NIARCHOS,[23] ORVILLE BRETTMAN,[24]
AND STAN WALKER[25]

[1] Department of Astronomy, Columbia University, 550 West 120th Street, New York, NY 10027; jop@astro.columbia.edu, abulafia@astro.columbia.edu
[2] Center for Backyard Astrophysics (Italy), Via Madonna de Loco, 47, 03023 Ceccano FR, Italy; gianmasi@fr.flashnet.it
[3] Rochester Institute of Technology, Department of Physics, 85 Lomb Memorial Drive, Rochester, NY 14623; mwrsps@rit.edu, smd5659@osfmail.rit.edu, tadsps@rit.edu

[4] King's University College, Department of Physics, 9125 50th Street, Edmonton, AB T5H 2M1, Canada; bmartin@kingsu.ab.ca
[5] Center for Backyard Astrophysics (Colorado), 14795 East Coachman Drive, Colorado Springs, CO 80908; ebeshore@pointsource.com
[6] Center for Backyard Astrophysics (East), 9517 Washington Avenue, Laurel, MD 20723; dskillman@home.com
[7] Joint Astronomy Centre, University Park, 660 North A`ohōkū Place, Hilo, HI 96720; j.kemp@jach.hawaii.edu
[8] Visiting Astronomer, Cerro Tololo Interamerican Observatory, National Optical Astronomy Observatories, which is operated by the Association of Universities for Research in Astronomy, Inc. (AURA) under cooperative agreement with the National Science Foundation
[9] Center for Backyard Astrophysics (Belgium), Walhostraat 1A, B–3401 Landen, Belgium; tonny.vanmunster@advalvas.be
[10] Center for Backyard Astrophysics (Nelson), 8 Regent Lane, Richmond, Nelson, New Zealand; reamarsh@ihug.co.nz
[11] Center for Backyard Astrophysics (Blenheim), 83 Vintage Lane, RD 3, Blenheim, New Zealand; wallen@voyager.co.nz
[12] United States Naval Observatory, Flagstaff Station, Post Office Box 1149, Flagstaff, AZ 86002; aah@nofs.navy.mil
[13] Center for Backyard Astrophysics (Indiana), 2507 County Road 60, Auburn, IN 46706; starkey@fwi.com
[14] Center for Backyard Astrophysics (Utah), 4175 East Red Cliffs Drive, Kanab, UT 84741; jfoote@scopecraft.com
[15] Center for Backyard Astrophysics (Finland), Vertaalantie 449, Nyrölä, Finland; arto.oksanen@jklsirius.fi
[16] Center for Backyard Astrophysics (Concord), 1730 Helix Court, Concord, CA 94518; lcoo@yahoo.com
[17] Center for Backyard Astrophysics (Flagstaff), Braeside Observatory, Post Office Box 906, Flagstaff, AZ 86002; captain@asu.edu
[18] Center for Backyard Astrophysics (Hamburg), Himmelsmoor 18, D–22397 Hamburg–Duvenstedt, Germany; husar_d@compuserve.com
[19] Nicholas Copernicus Observatory, Kravi Hora 2, Brno 616 00, Czech Republic; novak@hvezdarna.cz
[20] Arkansas Tech University, Department of Physical Science, 1701 North Boulder Avenue, Russellville, AR 72801; jeff.robertson@atu.edu, tutsky@yahoo.com
[21] Center for Backyard Astrophysics (New Mexico), 1688 Cross Bow Circle, Clovis, NM 88101; krajcit@3lefties.com
[22] Crimean Astrophysical Observatory, P/O Nauchny, 334413 Crimea, Ukraine; pavlenko@crao.crimea.ua
[23] University of Athens, Department of Astrophysics, Astronomy, and Mechanics, Panepistimipolis, GR–157 84, Zografos, Athens, Greece; pniarcho@cc.uoa.gr
[24] Center for Backyard Astrophysics (Huntley), 13915 Hemmingsen Road, Huntley, IL 60142; rivendell.astro@worldnet.att.net
[25] Center for Backyard Astrophysics (Waiharara), Wharemaru Observatory, Post Office Box 13, Awanui 0552, New Zealand; astroman@voyager.co.nz

**ABSTRACT**

We report the results of a worldwide campaign to observe WZ Sagittae during its 2001 superoutburst. After a 23-year slumber at $V$=15.5, the star rose within 2 days to a peak brightness of 8.2, and showed a main eruption lasting 25 days. The return to quiescence was punctuated by 12 small eruptions, of ~1 mag amplitude and 2 day recurrence time; these "echo outbursts" are of uncertain origin, but somewhat resemble the normal outbursts of dwarf novae. After 52 days, the star began a slow decline to quiescence.

Periodic waves in the light curve closely followed the pattern seen in the 1978 superoutburst: a strong orbital signal dominated the first 12 days, followed by a powerful *common superhump* at 0.05721(5) d, 0.92(8)% longer than $P_{orb}$. The latter endured for at least 90 days, although probably mutating into a "late" superhump with a slightly longer mean period [0.05736(5) d]. The superhump appeared to follow familiar rules for such phenomena in dwarf novae, with components given by linear combinations of two basic frequencies: the orbital frequency $\omega_o$ and an unseen low frequency $\Omega$, believed to represent the accretion disk's apsidal precession. Long time series reveal an intricate fine structure, with ~20 incommensurate frequencies. Essentially all components occurred at a frequency $n\omega_o-m\Omega$, with $m$=1, ..., $n$. But during its first week, the common superhump showed primary components at $n\omega_o-\Omega$, for $n$=1, 2, 3, 4, 5, 6, 7, 8, 9 (i.e., $m$=1 consistently); a month later, the dominant power shifted to components with $m=n-1$. This may arise from a shift in the disk's spiral-arm pattern, likely to be the underlying cause of superhumps.

The great majority of frequency components are red-shifted from the harmonics of $\omega_o$, consistent with the hypothesis of apsidal advance (prograde precession). But a component at 35.42 c/day suggests the possibility of a retrograde precession at a different rate, probably $N$=0.13±0.02 c/day.

The eclipses permit measuring the location and brightness of the mass-transfer hot spot. The disk must be very eccentric and nearly as large as the white dwarf's Roche lobe. The hot-spot luminosity exceeds its quiescent value by a factor of up to 60. This indicates that enhanced mass transfer from the secondary plays a major role in the eruption.

## 1. INTRODUCTION

WZ Sagittae is perhaps the world's most famous dwarf nova. Reaching magnitude 8 in eruption, it is the brightest of all dwarf novae; and the high binary inclination of 70–80° produces eclipses in its light curve, giving clues to the distribution of light over the various components of the binary system. The classic studies of the 1960s (Krzeminski 1962; Kraft, Matthews, & Greenstein 1962; Krzeminski & Kraft 1964; Krzeminski & Smak 1971) in quiescence established most of the basic binary parameters: an orbital period of 82 minutes, a secondary of very low mass, bright double-peaked emission lines from an accretion disk, and a very low luminosity (since about half of the visual luminosity comes from a mere white dwarf). All of these properties have been studied in much greater detail in the subsequent thirty years; an excellent review is given by Smak (1993). The very long outburst recurrence time of ~30 years has also anointed WZ Sge as the prototype of its own class of variable stars: dwarf novae of great faintness in quiescence ($M_V$=10–12), with very rare and long-lived eruptions (Bailey 1979; O'Donoghue et al. 1991; Kato, Sekine, & Hirata 2001).

Coverage of the 1913 and 1946 eruptions came primarily from visual observers. From the amplitude and recurrence time of the eruptions, WZ Sge was usually classified as a "recurrent nova" (although this term did not then imply a luminosity near the Eddington limit, as it does today). Krzeminski's discovery of the binary period catapulted the star to prominence, and thus the 1978 eruption was observed in much more detail, despite the unfavorable seasonal timing (December). The spectrum, the presence of flickering, and the periodic humps all attested to WZ Sge's proper classification as a dwarf nova [Patterson et al. 1978; but the recognition of low luminosity and classification as a dwarf nova were actually made much earlier, essentially by McLaughlin (1953) and Greenstein (1957)]. Studies of the 1978 eruption were published by Ortolani et al. (1980), Patterson et al. (1981, hereafter P81), and Mattei (1980).

In July 2001, WZ Sagittae erupted again into superoutburst. The very favorable seasonal timing enabled Earthlings to obtain long nightly light curves, and the star's celebrity status made it easy to motivate observers around the world. The result was the most thoroughly watched dwarf-nova eruption in history. This paper reports our photometric coverage, and especially our study of periodic signals, during and after eruption.

## 2. PHOTOMETRY AND THE ERUPTION LIGHT CURVE

WZ Sge was discovered bright by T. Ohshima (reported by Ishioka et al. 2001) on 23 July 2001, and confirmed within a few hours by astronomers at Kyoto University. As night fell progressively westward around the world, many telescopes turned to WZ Sge and began observing campaigns[1]. We report here an extensive campaign carried out by telescopes of the Center for Backyard Astrophysics, a network spread in longitude and designed to study periodic phenomena in variable stars (Skillman 1993). In all we accumulated 1220 hours over 325 nightly observations, distributed as given in the log of Table 1.

[1] Fascinating accounts and many, many light curves can be found at: http://www.kusastro.kyoto-u.ac.jp/vsnet/DNe/wzsge01.html

The data consisted of differential photometry with respect to a nearby comparison star, usually GSC 1621:1830 ($V$=8.74, $B–V$=0.17) and usually in unfiltered light. For most telescope/camera combinations, this implied an effective wavelength in the range 6000–6300 Å. We formed long time series by using overlaps to splice the individual runs, and calibrated individual nights with snapshot observations through standard *UBVRI* filters. The uncertainty in absolute calibration was typically ~0.15 mag, but the instrumental (differential) magnitudes were reproducible within 0.05 mag. After WZ Sge faded below $V$~12, we switched to a fainter star 1' south ($V$=11.75, $B–V$=0.19). Accurate photometry in the field has been reported by Henden & Landolt (2001).

WZ Sge has a close companion 10.9" to the west, which can present problems for photometry. This star has $V$=13.88, $V–R$=0.77, which is troublesome for our runs in unfiltered ("pink") light. We normally used small apertures to cleanly exclude the companion; in very bad seeing, we included the companion and corrected for the extra contamination. This was easy when the star was bright, but became difficult when WZ Sge faded below $V$~13 — a difficulty aggravated by the heterogeneity of our data. The effect of inclusion is mainly to degrade signal-to-background; fortunately it has no strong effect on our main program, the study of periodic terms in the light curve.

Unfiltered light also suffers from differential extinction (the blue CV is fainter at large airmass than the redder comparison star). In some cases we attempted to correct for this, but usually we discarded such data obtained at large airmass.

The upper frame of Figure 1 shows the eruption light curve, with a very rapid rise on JD 2452113.9 and a 24-day decline at an average rate of 0.10 mag/day, followed by a sharp 2-mag drop lasting 3 days. On JD 2452143 began the first of 12 remarkable short outbursts, each rising at a rate of 4 mag/day, lasting ~1 day, and falling at a rate of 0.70±0.13 mag/day. The lower frame shows a magnified view of this interesting region.

Most of the magnitudes in Figure 1 and Table 1 are averages over 2–6 hour runs, hence unaffected by variations on orbital and shorter timescales. We used our $V$ photometry to calibrate the unfiltered runs, which is fairly easy since the comparison star is of similar color, and WZ Sge shows little color variation (with $B–V$ and $V–R$ ≈ –0.03±0.10 through most of the eruption). WZ Sge was also watched very closely by visual observers, with results deposited in the AAVSO and VSNET archives. We measured the offset between these magnitudes and our CCD $V$, and used the former (averaging ~5 of them) to supplement the latter. This fills out the eruption light curve with essentially no gaps. The heterogeneity of conversion (true $V$, $V$ from unfiltered, $V$ from visual) increased the systematic total uncertainty in absolute calibration to ~0.15 mag. Random errors are much smaller, ~0.02–0.03 mag in the time series which are the

---

http://www.bellatrixobservatory.org/wzsge.htm
http://www.aavso.org/wzsge.stm
http://www.lunarpages.com/cbabelgium/wzsge_aug_2001.htm
http://www.kingsu.ab.ca/~brian/astro/cba_alta/data_archive/wzsge/wzsge.html

heart of this study.

In the analysis below, we frequently use *intensity* units, to make more transparent the actual changes in signal strength, eclipse depth, etc. We used the instrumental delta-magnitude scale with the primary comparison star set = 1000 counts. Thus our unit of intensity is approximately mJy (1 count = 1.2 mJy assuming a flat spectrum). This differed slightly from the *V* scale, because WZ Sge's continuum slope became redder during the final decline (thus increasing counts in unfiltered light).

Our photometry began on JD 2452114, the second day of eruption. We shall call this "day 14" of the eruption, and refer to dates as HJD–2452100. We follow convention and use the quiescent hot-spot eclipse as the zero-marker of orbital phase $\varphi$, with the most recently published ephemeris (Patterson et al. 1998a, hereafter P98a):

$$\text{Mid-eclipse} = \text{HJD } 2437547.72840(8) + 0.0566878460(3)\, E. \qquad (1)$$

Actual inferior conjunction of the secondary occurs 0.043±0.003 cycles earlier (Spruit & Rutten 1998, Steeghs et al. 2001, Krzeminski & Smak 1971). The orbital frequency is 17.640 c/day, a useful reference in the frequency forest that lies ahead!

## 3. LIGHT CURVES AND PERIODS DURING THE MAIN OUTBURST

Throughout the eruption, WZ Sge showed periodic signals in the vicinity of $P_{orb}$. To measure these periods accurately, we formed long time series over the intervals corresponding to the apparent stages of evolution. The light curves were found to be frequently *multiperiodic*, with the components beating together over 5–6 days. We therefore tended to select segments of 5–6 days for analysis. This gives adequate frequency resolution, and insures that synchronous summations will be minimally contaminated by the unwanted signal (since it sweeps through all phases in this interval). In addition, we usually subtracted the unwanted signal prior to summation; with this extra protection, we could adequately separate the components.

Because we concentrate on periodic signals, our standard procedure was to remove the mean and trend from each night's light curve. This removes unwanted power from all low frequencies. Some of the light curves displayed here have been "pre-whitened" in this manner. The estimated *V* magnitudes are given in Table 1, so purists can recover the original data.

WZ Sge shows occasional sharp dips in the light curve, absorption/occultation events associated with the binary orbit. The most obvious are associated with the bright-spot eclipse, though there are others of unclear origin. For periodic analyses, these are big trouble! So we removed sharp dips prior to calculating power spectra.

Finally, because of the need for variable-star jargon in this paper, we present in Appendix A a guide to *hump zoology* in CVs. Mercifully, WZ Sge does not display *all* these types of photometric waves; but it does display many, and this should help the reader navigate through the complexities.

### *3.1 DAY 14–25: THE OUTBURST ORBITAL HUMP*

From the beginning of our photometry, WZ Sge flashed a powerful signal at or very near $P_{orb}$. A sample light curve is shown in the top frame of Figure 2, while the lower frame shows the evolution of the nightly orbital light curve (in intensity units, with the mean intensity removed from each night) over the first 8 days of outburst. Obviously the orbital signal falls drastically in amplitude during this interval. The power spectrum of the 8-day light curve is shown in the upper frame of Figure 3. The strong signals are harmonics of a fundamental at 17.649±0.006 c/day, essentially consistent with the orbital frequency. This phenomenon was first identified in the early coverage of the 1978 outburst, and appears to be a standard signature of WZ Sge-type dwarf novae. We call it the *outburst orbital hump*[2].

Mean waveforms in the early and late parts of the signal's 12-day tenure are shown in the lower frames of Figure 3. Comparison shows the decline in amplitude (much more prominent in intensity units), and also a slight change in waveform. Light maximum drifted in orbital phase from $\varphi$=0.60 to 0.51 (each ±0.02). Does this mean that the period is not exactly $P_{orb}$? Yes, that's possible. But the other light-curve extrema are more stable, and the best value of the hump frequency in the power spectrum is essentially consistent with $\omega_{orb}$. This drift may signify a small change in waveform only, not a period truly different from $P_{orb}$ (but see Ishioka et al. 2002 for a contrary view).

We subtracted the dominant signals and studied the residual time series, to search for weaker periodic signals. The (incoherent) sum of the 4-day early/middle/late power spectra is shown in Figure 4, indicating an apparent signal at 17.52±0.03 c/d. This was a surprise, possibly an ancestor of the common superhump which developed strongly on day 26. The measured full amplitude appeared to decline from 0.070 to 0.022 mag over the 12-day interval; however, the phase of the signal wandered significantly (at least 10 times faster than the common superhump soon to come), so the amplitude measurement is not reliable.

There also appears to be a signal at 35.41±0.03 c/d, but no further details could be wrung out of this weak and uncertain detection. As we shall see below, this frequency popped up on other occasions during the outburst[3].

### *3.2 DAY 26–37: COMMON SUPERHUMPS*

[2] The suggestion of an anonymous referee several years ago. We like to reserve the term "superhump" for signals at periods displaced from $P_{orb}$. "OOH" is a useful shorthand, descriptive of a common emotional state when first observing these powerful waves in freshly erupted, long-dormant dwarf novae.

[3] In general we are wary of signals that appear only after more powerful signals are subtracted, since they can be mimicked by amplitude or phase changes. On these grounds, both signals are merely candidates in this interval; apparition in other time intervals makes the 35.41 c/d signal more secure.

On day 26, another periodic wave increased rapidly in amplitude: the *common superhump*, a feature of all SU UMa-type dwarf novae in superoutburst. During the common-superhump era, and unlike the OOH era, the light curve changed significantly from night to night. Mean orbital light curves are shown in Figure 5, with each frame averaging over 4–8 consecutive binary orbits. The most obvious feature is the sharp eclipse now (transiently) evident near $\varphi=0$, appearing on days 31, 32, 36, and 37.

Data obtained during the first week (day 27–33) are displayed in Figure 6. The upper frame shows the light curve of day 27, by which time the superhump was well-formed. Note some very narrow dips in the light curve; these occur at a repeatable orbital phase (0.695±0.008). The middle frames show the power spectrum[4] during that week. The superhump is very strong (0.14 mag full amplitude, rising to a power of 1030 at 17.49 c/d) and has a relatively simple fine structure. In the standard interpretation of superhumps, the dominant wave occurs at a frequency $\omega_o-\Omega$, where $\omega_o\equiv\omega_{orb}$ and $\Omega$ is the unseen precessional frequency. With this terminology, the detected signals were $\omega_o-\Omega$, $\omega_o$, $2\omega_o-2\Omega$, $2\omega_o-\Omega$, and $3\omega_o-\Omega$. The lowest frames show the mean superhump and orbital light curves. The former is essentially the familiar fast-rise–slow-decay waveform of common superhumps. The weak orbital signal is more difficult to assess. The sharp dip at $\varphi=0$ is obviously the eclipse; the broader feature flanking it could be real, but could also arise from imperfect removal of the powerful superhump.

Light curves during the second week appear similar; but the power spectrum, seen in the upper and middle frames of Figure 7, shows great changes. The orbital signal is stronger, and signals at higher frequency are more complex. Detections occurred at $n(\omega_o-\Omega)$, with $n$=1, 2, 3, 4; and $n\omega_o-\Omega$, with $n$=2, 3, 4, 5, 6, 7, 8, 9. Other components occurred at $3\omega_o-2\Omega$, $4\omega_o-3\Omega$, and $5\omega_o-3\Omega$, as well as an uncertain detection at 70.68 c/day, i.e. $4\omega_o$+0.12 c/day. This is "the spectrum of the common superhump" — during its second week.

The orbital and superhump waveforms are seen in lowest frame of Figure 7. The orbital wave somewhat resembles that of quiescence. A sharp dip occurs at $\varphi=0.001\pm0.002$, the correct time to eclipse the bright spot (where the mass-transfer stream impacts the edge of the disk); but the eclipse feature is wider, with a full-duration at half-depth of 350±25 s, compared to 164±10 s in quiescence (Robinson et al. 1978, hereafter RNP).

### *3.3 O–C ANALYSIS*

---

[4] The sharp eclipse feature severely contaminates periodic analysis, so we always removed it prior to analysis. We also sometimes chose to "clean" the power spectra by successive removal of the strongest signals, although it was often unnecessary since our nightly coverage was typically very long (this technique is principally used to remove aliases).

O–C analysis is usually a more accurate method of tracking periodic waves in dwarf novae, because it can use sharp features in the light curve, and because it is not confounded by changes in amplitude or mean light level[5]. We timed maxima and minima in the light curve, and show the results in Table 2. We list individual timings during episodes when the measurements were easy, and condense them into averages of three consecutive timings when there were difficulties due to low amplitude, flickering, and confusion from nearby events (often eclipses) in the light curve. We limit this discussion to the main outburst, days 14–38. Figure 8 shows the O–C departures of maxima from the eclipse ephemeris. The trend is nearly flat (declining slightly from φ=0.60 to 0.51) during the era of OOH dominance, ending at about day 25. During day 26–37, it switches to a new slope, with $P$=0.05719(5) d. The exact value is somewhat unreliable; the light curves were sometimes disturbed by apparent sharp absorption dips near maximum light, producing systematic residuals in the O–C (visible on close inspection of Figure 8). The power spectrum yields 0.05724(5) d, so we adopt 0.05721(5) d as a mean period. In Section 7 we shall return to the issue of periods and period changes.

## 4. THE DIP: DAYS 39–42

During days 38–40, the star faded rapidly (at 0.7 mag/day) to a minimum at $V$~12.8, where it remained for 3 days. Figure 1 shows this first fading event. The nightly light curves were complex, with both orbital and superhump signals. Since these signals drift by only ~0.1 cycles in the span of one night, we can represent the behavior fairly well with nightly synchronous summations at $P_{orb}$. These are shown in Figure 9. As the star fades (days 38–39), the light curve is complex and double-humped, with a bright-spot eclipse becoming evident. The deep minimum (days 40–41) shows a pronounced eclipse and "orbital hump" — indeed, these waveforms resemble the orbital light curve at quiescence (RNP).

On day 42, the waveform was complex, presumably because orbit and superhump competed strongly. The star stayed faint for ~4 hours ("day 42a"), then rose sharply in brightness ("day 42b"). The hump amplitude and eclipse depth stayed approximately constant in intensity units, indicating that the source of the rising light was neither the source of the hump, nor the object eclipsed.

## 5. THE ECHO OUTBURSTS: DAYS 43–65

Day 43 saw the first of 12 small and fast outbursts in the light curve. These are an occasional feature of the decay from superoutburst in WZ Sge stars. Since they follow directly after the main eruption, they have been called "echo" outbursts, and we shall use that term here. The most thoroughly studied such star is EG Cancri (Matsumoto et al. 1998; Patterson et al. 1998b, hereafter P98b).

The light curve of the entire episode of echo-outbursts is shown in Figure 1, and a 2-night light curve is shown in Figure 10. In section 9 below we will discuss the origin of these outbursts. Here we consider their effects — which are mighty pernicious — on the detection of periodic signals. The power in the outbursts is enormous and will leak all over the power spectrum if no measures are taken to remove it; and there is no sure-fire technique of removal. In the present case, we noticed that the amplitude of the periodic signal becomes much lower when the star is bright, and amplitude changes bedevil Fourier methods. We converted the delta-magnitude time series to intensity units, and then treated these 2-day outbursts as unwanted extra light with no intrinsic modulations near $\omega_o$. This assumption is probably not correct in detail; but after subtracting that unwanted light, we did obtain time series similar in properties to the time series away from outbursts. In other words, the results were consistent with the hypothesis that the periodic signals are long-enduring phenomena which are merely diluted by an extra unmodulated source which is the outburst light. This was also found for the echo-outbursts of EG Cnc, where the superhumps proceeded through the whole episode with no obvious change in amplitude or period (see Figure 5 of P98b). So we adopted that assumption.

[5] These issues famously trouble Fourier techniques. However, the latter are far superior in demonstrating the *existence* of a periodic signal, since they permit examination of the noise in frequency space.

Even with this improvement, the power spectra in this interval are complex. There appear to be several components near each harmonic of $\omega_o$, and roughly orbital absorption events which show a complex (not quite decipherable) pattern of disappearance and reappearance. Finally the strength of components and their spacing in frequency changed significantly over this 22-day interval. So we divided the interval into halves, and limit presentation to those components which we judge to be reliable.

The cleaned power spectrum of the first half is shown in the upper row of Figure 11. The signals detected are consistent with: $n(\omega_o–\Omega)$, for $n$=1, 3, 4; $n\omega_o$ for $n$=1, 2, 4; and components at $2\omega_o–\Omega$, $3\omega_o–2\Omega$, $3\omega_o–\Omega$, and $4\omega_o–3\Omega$. (See the caption for the special treatment of the region near $4\omega_o$.) Averaged over these detections of harmonics and fine structure, the best estimate of $\Omega$ is 0.238±0.006 c/day. The cleaned power spectrum of the second half is shown in the middle row of Figure 11. In addition to the smallest integer multiples of $\omega_o$ and $\omega_o–\Omega$, the signals detected include $2\omega_o–\Omega$, $3\omega_o–2\Omega$, $3\omega_o–\Omega$, $4\omega_o–3\Omega$, and $4\omega_o–\Omega$. Finally there are blue-shifted components near 17.84 and 35.43 c/d, which we discuss in §7.3 below.

The bottom row of Figure 11 shows the cleaned power spectrum of the entire 22-day segment. This improves frequency resolution, and better establishes the reality of the blue-shifted components. But the higher harmonics are not shown, because changes over the 22 days introduce confusion (compare the higher harmonics in the upper and middle rows of Figure 11).

We also experimented with finer slices of the 22-day segment. The general results were the same: the strongest components near each harmonic occurred at $2\omega_o–\Omega$, $3\omega_o–2\Omega$, and $4\omega_o–3\Omega$. In contrast, the first week of the common superhump showed primarily a simple signal at $\omega_o–\Omega$, while the second week showed considerable structure but dominated by the $n\omega_o–\Omega$ components.

## 6. FINAL DECLINE

On day 67, WZ Sge faded rapidly from its final echo, and resumed a slow decline towards quiescence. The decline rate averaged 0.02 mag/day over the next 50 days. The mean nightly orbital light curves are shown in Figures 12–14, with the orbit resolved into (usually) 100 independent phase bins. Inspection of these figures shows that the pattern of variation repeats with a cycle of ~5 days (corresponding to the beat period of orbit and superhump).

We divided the decline into 10–11 day segments, which gave good frequency resolution without excessive smearing of the signals. Relevant portions of the power spectra are shown in Figure 15 for days 68–79, 80–91, and 89–99. The general pattern from previous weeks persists — but with some differences. The powerful $\omega_o$–$\Omega$ superhump is simple (little or no power at harmonics), and the strongest of the other components seem to occur at $n\omega_o-(n-1)\Omega$.

Synchronous summations at the orbital and superhump period are given in Figure 16. The superhump evolution is simple: the waveform is closely sinusoidal, with an amplitude decreasing throughout. The orbital hump and the eclipse depth also decrease gradually with time.

## 7. PERIODS OF PHOTOMETRIC WAVES

Since WZ Sge displays many noncommensurate signals of variable amplitude, accurate measure of the periods is difficult. The O–C diagrams are often confounded by interference from nearby frequencies, and/or by absorption events which are difficult to identify unambiguously. We found that the safest procedure was to limit analysis to segments where orbit and superhump could be separated, preferably with an integer number of beat cycles elapsing. We also measured the periods generally from the power spectrum, since the O–C diagrams were problematic and somewhat subjective.

### *7.1 THE OUTBURST ORBITAL HUMP*

The properties of the OOH were studied in Sec. 3.1. The period was 0.056666(20) d, about 1$\sigma$ from $P_{orb}$. As stated above in 3.1, we are inclined to consider this to be $P_{orb}$, although it is certainly possible that the true period is slightly shorter. The brevity of the 12-day baseline, and the slight change in waveform, make this issue not quite decidable. In intensity units, the signal declined by 36% per night (see Figure 2).

### *7.2 THE SUPERHUMP(S)*

As is typical for common superhumps, these waves increased very suddenly in amplitude — despite the 12-day wait for their appearance. The mean waveforms (Figures 6 and 7) were also typical of common superhumps, but the nightly waveforms were highly variable (Figure 5). Why? The reason is that the superhump was disturbed by phenomena occurring on the 5–6 day beat cycle — the drifting phase of the orbital wave, but more importantly, sharp dips which appear to be absorption events. This is probably the main reason that the O–C diagram of Figure 8 showed a troublesome series of linear segments with discrete jumps.

Table 3 gives the measured periods and amplitudes over the selected intervals, after removing the orbital signal. The full amplitude declined from 0.18 to 0.06 mag during the main outburst, and then declined again from 0.20 to 0.06 mag from day 68 to 109. During echo outbursts, the amplitudes were lower and more uncertain. The periods given by the power spectra are shown in the upper frame of Figure 17, and show a slight increase over the 80-day baseline.

We subtracted the orbital waves in each (6–11 day) segment, and then averaged superhump maxima during each night. The O–C diagram of the resultant superhump maxima is shown in the lower frame of Figure 17, relative to a test period of 0.05727 d. The slopes in the O–C agree with the periods deduced from the power spectrum, from 0.05721(6) d during the main outburst to 0.05736(5) d near day 100.

Superhumps in some dwarf novae show a rapid phase shift of ~0.5 cycles late in the decline, with little or no change in period. These are called *late* superhumps (Vogt 1983, Hessman et al. 1992). The phase shift is the property that defines the transition from common to late superhumps, so we studied the O–C diagram to look for that transition. Figure 17 does not clearly specify it. However, there *is* a rapid phase change as the main outburst ended on day 39, the phase *does* stabilize ~12 d later, and the superhumps certainly last a long time (improving their credentials for being considered "late"). The O–C diagram bears a fair resemblance to that of EG Cnc, one of the better-documented late superhumpers (compare Figure 17 with Figure 5 of P98b). So we will designate this last phase as "late", even though evidence for membership in that class is not conclusive.

Tick marks on the O–C diagram indicate the times of echo outbursts, demonstrating that the superhump clock is not affected by echoes.

### *7.3 BLUE-SHIFTED SIGNALS?*

The power spectra show two signals blue-shifted from the harmonics of $\omega_o$. The signal at 35.42±0.02 c/day is always weak, but is detected on several occasions, and we consider it a fair candidate as a *negative superhump*. Under a popular interpretation of such things, this implies a retrograde precession $N$=0.14±0.02 c/day.

A signal at 17.84±0.01 c/day also appears several times. This detection was fairly strong. It must be considered suspect, though, because of its displacement from $\omega_o$ by 0.20(1) c/day, consistent with $\Omega$. Strong amplitude modulation of the orbital signal at $\Omega$ will produce power at $\omega_o-\Omega$ and $\omega_o+\Omega$; the former would be lost in the powerful superhump, but the latter could account for the observed signal. We could reproduce the signal in simulations with amplitude changes alone; thus, in the absence of more evidence, we do not yet credit this as an independent signal in WZ Sge.

### *7.4 THE 1978 OUTBURST*

The periodic behavior in 1978 was very similar. The first 12 days of that eruption showed a ~0.2 mag variation at or very near $P_{orb}$, followed by a several-week superhump accompanied by a residual orbital signal (P81). This was identical to the 2001 behavior. The exact period of the 1978 superhump depends on the treatment of the contaminating orbital signal. Assuming quasi-sinusoidal signals (i.e. large duty cycle), P81 estimated 0.05714(4) d. O–C analysis is slightly more accurate but essentially assumes a short duty cycle, i.e. resolution of orbit and superhump in a single night's light curve, or ignoring contamination by the orbital signal. The latter analysis gave values of 0.05725 d (the same McDonald Observatory data, cited by Targan 1979), 0.05722 d (Targan 1979) and 0.05723 d (Bohusz & Udalski 1979), with a similar error. The difference is small, and not decidable in data severely limited by poor seasonal timing and with great intrinsic complexity (two signals plus dips). As a compromise we adopt 0.05720(5) d for the 1978 period, consistent with that of the 2001 common superhumps.

## 8. THE ECLIPSE AND THE ORBITAL WAVE

Eclipses can be a powerful diagnostic of the underlying structure in the binary. But the quiescent WZ Sge is inclined sufficiently to eclipse only the mass-transfer bright spot, not the white dwarf or inner disk (Krzeminski & Smak 1971). This implies less information in the eclipse waveform, but makes it easier to measure the brightness of the hot spot, which is essentially an instantaneous barometer of the mass-transfer rate.

P81 studied this issue for the 1978 outburst, and concluded that there was evidence for eclipse features timed exactly right to be hot-spot eclipses (Figure 3 of P81). The 2001 coverage makes this obvious. The eclipse first appeared on day 27, and persisted through late decline. However, a glance at the nightly light curves (Figures 5, 9, 12, 13, 14) shows that it drifts in and out of view with a period ~5 days. On this period, the hot spot must either disappear, or drift in location so it moves out of the eclipse shadow. In Section 12 we will address how this can occur with an eccentric disk. Here we present measures of the intensity of the hot spot, revealed by measures of the eclipse depth and the orbital wave height.

These are given in Figure 18. We have measured eclipse depths by examining the mean nightly orbital light curves, and selected the night of maximum depth in the ~5-day cycle. This is to facilitate the comparison with quiescence (some other effect, related to precession, reduces the depth on other nights — but is not present in quiescence). The eclipse is weakly seen on day 27, but then appears with great depth on days 32 and 37 (see Figure 5). The depth then declines smoothly through the rest of outburst. On the scale of Figure 18, the eclipse depth at quiescence is 0.4, so the hot-spot appears to be enhanced by a factor of ~60 in outburst.

Does the eclipse depth directly measure the intensity of the hot spot? Not quite. At quiescence, the wave associated with the hot spot is near a minimum when the eclipse occurs, so the eclipse depth underestimates $I_{spot}$. The same is true in outburst, because the orbital wave is similar. An alternative is to use the height of the orbital wave. However, this too underestimates $I_{spot}$, because the spot is somewhat visible on the back side of the disk (or to put it another way, the wave amplitude measures only the asymmetric part of the spot's radiation pattern). Anyway, we measured the orbital waves from each segment analyzed (after subtracting the more powerful superhump) and found the amplitudes given in the right frame of Figure 18. The pattern is similar: an amplitude declining from 13 to 0.4 (the quiescent value) during the outburst.

## 9. ECHOES

These hot-spot eclipses establish that mass-transfer is greatly enhanced during superoutburst. This may settle the debate over the origin of echo outbursts in dwarf novae.

Several ideas have been proposed to explain echoes. The observational evidence (timescale, spectrum, behavior of periodic signals) establishes fairly well that these are some variant on "normal" dwarf-nova outbursts. The question is, why should they occur so frequently after superoutburst, and then die out altogether? The most developed suggestions are those of Osaki et al. (1997, 2001) and Hameury et al. (2000). Osaki et al. proposed that after the main superoutburst, enough matter resides in the outer torus to drive subsequent thermal instabilities — but the emptying torus and the declining viscosity become quickly insufficient (mainly because of the viscosity decline). The model of Hameury et al. also invokes thermal instabilities, but relies on the enhanced mass-transfer to drive the binary through an $\dot{M}_2$ regime appropriate for producing the short eruptions.

Since the depth of the hot-spot eclipse is a pretty good barometer of $\dot{M}_2$, we can use Figure 18 to test the latter theory. The echo era is indicated in the figure, as well as the eclipse depth expected for $\dot{M}_2 \sim 10^{16}$ g/s. During the echo era $\dot{M}_2$ appears to be in the range (0.7–2.0)x$10^{16}$ g/s, a range thought to be characteristic of dwarf novae (Hameury et al. 1998, Cannizzo et al. 1988, Osaki 1996). Observation and theory agree fairly well, suggesting that enhanced mass transfer is the principal cause of echo outbursts.

Nevertheless, it is curious that all echoes are of about equal amplitude and recurrence time, followed by a complete cessation. This was also true for EG Cnc. But it does not appear to be naturally explained by either of the models discussed; further work on this point is very desirable.

## 10. MASSES AND ENERGETICS

A thorough analysis of energetics in WZ Sge was presented by Smak (1993), based in part on the optical-UV fluxes revealed in the 1978 outburst. Comparison of the visual light curve and early reports of UV/X-ray fluxes in 2001 (Kuulkers et al. 2002) with that of 1978 indicate that the recent outburst was a pretty faithful reprise of 1978. We have repeated Smak's analysis and found similar results, with one important exception.

Namely, the white dwarf mass $M_1$. Smak estimated $M_1$=0.45 $M_\odot$, based principally on an assumed detection of the white dwarf's orbital motion (Gilliland, Kemper, & Suntzeff 1986). But that estimate is certainly too low, for reasons discussed in several recent studies [in quiescence: Patterson 1998 (hereafter P98), P98a, Spruit and Rutten 1998 (hereafter SR); in outburst: Steeghs et al. 2001). These latter estimates have their own problems, however. The 0.8±0.2 $M_\odot$ estimate of P98 used the 1440 km/s separation of emission-line peaks, which may not reflect the true Keplerian motion in the disk. And the 1.2±0.25 $M_\odot$ estimate of SR relied on an assumed detection of the white dwarf's motion, even though its phase was discrepant by 50°. The implied SR secondary-star mass of 0.09 $M_\odot$ is definitely ruled out by luminosity constraints.[6] (This problem affects mainly the estimate of $M_2$, though; the need for a high white-dwarf mass remains.)

Three other important numbers have recently become available. Thorstensen (2001, private communication) has obtained a trigonometric parallax indicating a distance of 43±8 pc; Sion et al. (1995) have measured the temperature of the white dwarf in quiescence to be 15000 K; and Steeghs et al. (2001) have determined $M_1$>0.77 $M_\odot$ from a measurement of $K_2$ in eruption (using the "chromospheric" emission of the secondary). Since white dwarfs obey a mass-radius relation, we can express these constraints as in Figure 19. The curve corresponds to the mass-$M_V$ relation for white dwarfs at 15000 K (Wood 1992, as applied in Figure 2 of Liebert et al. 1997), assuming the white dwarf to have $V$=16.2±0.3 at quiescence (our estimate). If the white dwarf is uniformly luminous, then it must lie on the curve, with $M_1$=1.23±0.15 $M_\odot$. If only part of the white dwarf surface radiates[7], then a slightly lower $M_1$ is permitted. The black region is consistent with all constraints, and we shall adopt an estimate of $M_1$=1.0±0.2 $M_\odot$. It may be of interest that this is just the range (formally $1.03^{+0.23}_{-0.20}$) permitted by the rotational velocity estimated from the UV line profiles ($1200^{+300}_{-400}$ km/s, Cheng et al. 1997), under the assumption that the latter is associated with $P_{rot}$=28 s (Patterson 1980).

The outburst energetics are straightforward. Most of the energy is radiated in the 1000–9000 Å window, with a flux distribution flat in frequency ($F_\nu \propto \nu^0$), as typically found for accretion disks. Integration under this curve yields a total received flux. (Alternatively, a bolometric correction of −1.8 mag can be applied, since that is a suitable correction for the ~20000 K temperature appropriate to this slope.) We also correct for the disfavored edge-on view of the disk, with $i$=75° suggesting a correction of 2.8× (Smak 1993). Then the total energy radiated over the 25-day main outburst is

$$E = 4.6 \times 10^{40} \text{ erg } (d_{43})^2,$$

where $d_{43}$ is the distance in units of 43 pc. About 20% more is radiated over the next 100 days. This radiation from a disk, neglecting boundary layer emission, implies a total mass accreted

$$\Delta M = 4 \times 10^{23} \text{ g } (d_{43})^2 \, (M_1/M_\odot)^{-1.8}.$$

---

[6] WZ Sge in quiescence has $K$=13.3 and shows no spectral or photometric features attributable to the secondary in this (or any other) wavelength regime (Dhillon et al. 2000, Ciardi et al. 1998, Littlefair et al. 2000). Thus the secondary has $K$>15.3, or $M_K$>12.2. This marks it as a brown dwarf; the end of the main sequence occurs at $M_K$~11, and 0.09 $M_\odot$ stars have $M_K$~9.5 (Henry & McCarthy 1993, Chabrier & Baraffe 2000, Figure 4 of Patterson 2001, Figure 3 of Baraffe et al. 1998).

[7] The point is potentially a crucial one, because we do not yet understand in detail the origin of the white-dwarf light. Greenstein (1957) first recognized the dominance of the white dwarf in the spectrum. Previous estimates of its temperature have been made: 12000–15000 (Krzeminski & Smak 1971), 10000–18000 (Patterson & Raymond 1985), and 12500 K (Sion 1991). For definiteness we use the most recent HST estimate (14900±250 K). But this temperature will need to be revised upward for a more realistic gravity (log $g$=8.3–8.8 rather than the 8.0 used to derive the temperature). That makes the luminous area smaller to accomodate the observed flux at a given distance, and hence moves the "whole photosphere" curve towards even larger $M_1$. Unless $M_1$ really is that large, we would have to conclude that the radiating surface is less than a full white-dwarf hemisphere.

This is a factor of ~4 less than estimated by Smak (1993); the difference arises from $M_1$.

The analysis at quiescence is more uncertain. The hot spot produces the orbitally modulated component, whose radiation is highly directional. At eclipse the radiation is directed ~80° away from the line of sight, which is probably a fairly representative direction (since it varies 0→180°). Adopting the eclipsed flux and a bolometric correction of –1.4 mag (the spot has quite blue colors; Krzeminski & Smak 1971), we estimate the spot to have a luminosity

$$L_{\text{hs}} = 3.5 \times 10^{30} \text{ erg/s } (d_{43})^2.$$

Assuming the hot spot to be formed at $R_{\text{disk}} = 0.4a$, this implies a mass transfer rate $\dot{M}_2$ given by

$$\dot{M}_2 = 1.0 \times 10^{15} \text{ g/s } (d_{43})^2 \, (M_1/M_\odot)^{-2/3},$$

or a total mass transferred of $10^{24}$ g over 30 years.

Despite the large change in the adopted $M_1$, Smak's conclusion is undisturbed: mass transfer at quiescence, accumulated over 30 years, appears sufficient to power the outburst. The *total energy budget* does not require any enhanced mass transfer during eruption.

## 11. THE PRECESSION CYCLE

The origin of superhumps was first identified by Whitehurst (1988), who described them as arising from an eccentric instability in the accretion disk. Perturbation by the orbiting secondary then leads to precession of the eccentric disk, and a strong periodicity at the lower precessional sideband $\omega_o - \Omega$, the frequency of tidal forcing. The precessional frequency $\Omega$ is itself generally unseen; there is no expectation of luminosity produced at that frequency. But in WZ Sge, the disk structures occulted by the grazing eclipse must depend on the apsidal orientation, so there *is* an explicit signature of precession phase apart from the superhump itself. We have examined all light curves for "time of deepest eclipse" in the 5-day cycle, and present these timings in Table 4.

The interval between events is a simple geometrical marker of the period of the 5-day cycle. In Figure 20 we show the variation of precession period with time during the outburst, based on a running average over three consecutive timings. Generally the period falls from ~5.5 to ~4.5 days, consistent with the periods independently deduced from the beating of orbit and superhump (Figure 17). This may be useful, because "deepest eclipse" is a more easily interpreted phase marker than "maximum superhump light".

## 12. ECLIPSES AND ECCENTRICITY

Eclipses provide an opportunity to measure the size and shape of the accretion disk. Alas, for WZ Sge, the eclipse of the disk proper is never clearly seen in the light curve, probably because the binary inclination is too low to place much of it in the secondary's shadow. But sharp eclipses of the mass-transfer hot spot are seen in many of our light curves. We measured those of adequate quality, and analyzed them with methods similar to those described by Smak (1996) and Hessman et al. (1992). Since most of our telescopes are quite small, the precision of our light curves is usually low; and we expect that data sets of higher quality will quickly supersede ours. Nevertheless, we present a preliminary study here, in the hope that the results are of interest.

### *12.1 DURING THE COMMON–SUPERHUMP ERA*

For the last half of the main outburst, common superhumps dominated the light curve. Measurement of days 31–39 revealed three effects which varied with the putative precession period of 5.4–6.0 d. One is the eclipse depth, shown in Figure 5. By measuring the moments of mid-ingress $t_i$ and mid-egress $t_e$, we also calculated the time of mid-eclipse $[(t_i + t_e) / 2]$ and the eclipse duration $(t_e - t_i)$. The absolute phasing of each quantity on the 5.7 d cycle (the beat of $P_{orb}$ and $P_{sh}$=0.05725 d) is shown in the left frames of Figure 21. Each evidently varies with precession phase. The zero-phase marker in the precession cycle was taken to be the time (day 34.135) when superhump maximum and orbital eclipse coincided.

The eclipse appears to be deepest at precession phase $\Phi$=0.69±0.04, latest at 0.71±0.05, and widest at 0.48±0.03. However, the lack of data during ~40% of every cycle (when the eclipses essentially disappear) implies that the detailed shape of these curves is not accurately specified; we fit sinusoids merely for simplicity.

### *12.2 DURING THE LATE–SUPERHUMP ERA*

Somewhat before day 55, a transition to "late" superhumps probably occurred. Eclipse measurements during the remaining ~10 days of echoes are difficult to make, so we restricted analysis to days 68–92; the mean superhump period of 0.05734 d (Figure 17) implies a putative precession period of 4.95 d. The same three quantities were found to vary with precession phase, as seen in the right panels of Figure 21. The eclipse is deepest at $\Phi$=0.87±0.03, latest at 0.88±0.04, and widest at 0.60±0.03.

Roughly speaking, the timing events in both precession eras are the same: the eclipses are deepest and latest at the same phase, which follows the phase of widest eclipse by 0.25±0.05 cycles. The absolute phasing is less secure. It appears to differ between common and late superhumps by 0.15 cycles; but the accuracy of the latter number depends on how accurately we know the superhump ephemeris, which is complex (Figure 17). We estimate the likely uncertainty as 0.07 cycles, hence regard this difference as a $2\sigma$ effect — significant, but not beyond doubt.

### *12.3 INTERPRETATION*

With data of relatively low precision, and with no white dwarf eclipse available, we did not attempt a formal solution for the disk dimensions. But a few remarks are warranted.

The geometry of the eclipse is shown in Figure 6 of RNP. The key feature is that the eclipse width is a measure of the chord on the secondary where the eclipse occurs, and should be greatest when the hot spot is closest to the secondary (largest $R_{disk}$). With prograde precession, the apastron of the disk should be maximally leading the secondary ~0.2 precession cycles later. That should produce a maximally late eclipse, as observed, and a deep eclipse (since the gas falls farther to the disk). These agree roughly with observation. We estimate an eccentricity $e \geq 0.3$ from the variations seen in Figure 21.

Better data of this type, and a more thorough analysis, could yield the shape and dimensions of the precessing disk [as Hessman et al. (1992) did for OY Car, and Rolfe et al. (2001) did for IY UMa]. This is especially required to test the tidal-instability theory discussed below, since the latter depends on the disk extending to the 3:1 resonance at $R_{disk} \sim 0.46a$.

## 13. THE OUTBURST ORBITAL HUMP

The orbital hump is a strong feature of the outburst. At least 15% of the entire optical-IR energy of the outburst is contained in this signal (by contrast, the common superhump is only ~3%). It appears within 1 day at high amplitude, maintains an essentially constant double-humped waveform, and then is replaced by a common superhump which develops in a normal manner. These properties (amplitude, waveform, period, development/decay timescales) of both signals appear very similar to those of the humps studied in the 1978 outburst (P81).

Three explanations for the OOH have been proposed:

(1) Heating of the secondary by light from the freshly erupted disk or hot white dwarf (Smak 1993).

(2) A burp of mass-transfer from the secondary, with a resultant hot-spot at the disk's outer edge (P81; Lasota, Hameury, & Hure 1995; Hameury, Lasota, & Hure 1997).

(3) A premature form of a superhump ("early superhump", Kato et al. 1996).

(1) is simple and plausible, but has two flaws which appear to be fatal. Maximum light is observed around $\varphi=0.60$, but the secondary reaches superior conjunction at $\varphi=0.46\pm0.02$, a serious discrepancy. Also, the observed waveform is double-humped, inconsistent with simple heating of the secondary.

(2) has interest because we have excellent evidence that $\dot{M}_2$ was enhanced by a factor of ~40 during day 31–37, and only declined to its quiescent value on the same timescale as the eruption light. Might $\dot{M}_2$ have been yet another factor of 10 higher in the first few days of the eruption, as required to accept the extreme form of (2)? Perhaps, but some excuse must be found for the absence of hot-spot eclipses with the familiar shape and phasing. It is quite incriminating that the evidence of hot-spot eclipses did not appear until the OOH *vanished*! Also, 30 years of mass-transfer is already sufficient to power the outburst; we ought not to be too eager to invite large quantities of extra matter over into the disk.

(3) has no deeply incriminating flaws but is too incompletely specified to evaluate. It does not (yet?) explain why the signal occurs at $P_{\rm orb}$, why its pattern of rise and fall differs so markedly and consistently from that of the common superhump, and why it exists at all.

Thus none of these explanations is quite satisfactory. A fourth possibility is suggested by theoretical studies of freshly erupted dwarf novae: the development of a strong *m*=2 spiral-arm structure at the beginning of an outburst, before the 3:1 eccentric resonance is strongly encountered (e.g. Whitehurst 1994, esp. his Figures 2 and 7; Simpson & Wood 1995; Truss, Murray, & Wynn 2000). This structure is fixed in the orbital frame, and therefore does not lead to any variable dissipation. But at high binary inclination, the aspect of the disk presented to distant observers in the inertial frame (us) varies with $P_{\rm orb}$. Of course, a deviation from axial symmetry also causes the photometric signal at $P_{\rm orb}$ in quiescence; since it manages a ~70% effect in quiescence (after subtracting non-accretion light), its cousin in outburst should be able to manage ~20%. The *m*=2 mode suggests a double-humped waveform, as observed.

This implies that the most highly inclined binaries should show the largest OOHs. The sparse data on this point appear to be consistent: large waves were seen in the two eclipsers, WZ Sge and possibly DV UMa (Patterson et al. 2000); the weakest wave was seen in EG Cnc, a low-inclination binary (attested by a failure to show an orbital signal at quiescence; P98b); and waves of intermediate strength were seen in HV Vir and AL Com, which do not eclipse but manage to sport an orbital wave at quiescence (Kato et al. 2001, 1996; Patterson et al. 1996).

Also supporting this idea is the spectroscopic observation of a 2-armed spiral in the Doppler tomograms of WZ Sge in the first few days of outburst (Steeghs et al. 2001, Baba et al. 2002). This seems altogether like a promising way to account for the OOH.

A simple analytic theory of this type (invoking a 2-armed spiral at the 2:1 resonance) has been recently proposed for WZ Sge by Osaki & Meyer (2002, hereafter OM). The OM theory starts as (nearly) all pure-disk theories start: the quiescent disk sits innocently as a ring at the Lubow–Shu or "circularization" radius $R_{\rm circ}$, when a sudden rise of disk viscosity triggers accretion. Some of the ring spirals in, and the rest spirals out, to conserve angular momentum. Assuming a steady-state disk structure, the outer edge of the disk increases in radius by a factor $(7/5)^2$. But $R_{\rm circ}/a$ is likely much bigger for binaries of very low *q*; the secondary hogs most of the angular momentum, so freshly transferred gas has more angular momentum and takes up residence in a bigger orbit. For our favored *q*=0.045, Table 2.1 and Eq. (2.18) of Warner (1995) imply $R_{\rm circ}$=0.38*a* — compared to 0.25*a* for a binary with *q*=0.2. Then when the disk expands by a factor $(7/5)^2$, both binaries will reach the 3:1 orbital resonance at $R_{\rm disk}$=0.46*a*, the binary of low *q* can charge right through it and reach the 2:1 resonance at $R_{\rm disk}$=0.63*a*. This leads to a strong tidal dissipation which releases energy (the early and brightest phase of the outburst) and drains angular momentum from the outer disk (terminating this phase pretty fast).

Though not without problems, a model of this type seems very attractive for the OOH. OM provide a lucid explanation for the one prominent feature not previously explained: the limitation of these powerful waves to binaries of low $q$.

## 14. THE ERUPTION

### *14.1 THEORIES*

The pioneering work of Hoshi (1979) established the physical basis — the onset of opacity as disk temperatures rise past 8000 K — for the modern understanding of dwarf-nova eruptions as thermal instabilities in the accretion disk. Whitehurst's (1988) study of disk dynamics revealed also a "tidal" instability in accretion disks, whereby disks become eccentric and subsequently precess under perturbation by the orbiting secondary. Many later studies have expanded our knowledge of these instabilities, and shown that they are central features of dwarf-nova eruptions.

However, we have not yet securely learned how these instabilities are related. The most obvious signs of this are: some dwarf novae do not superhump (the U Gem stars); some superhumpers are not dwarf novae (the permanent superhumpers); and dwarf-nova superhumps occur only in superoutbursts. The first two are easily understood: superhumps require an adequately low mass ratio ($q$<0.3), and should exist as long as high viscosity keeps the disk radius adequately high. These points are part of all the currently viable theories. The most popular and elegant way to understand the third point, as well, is the "thermal-tidal instability" (TTI) of Osaki (1989, 1996). In this theory a normal outburst occurs from the thermal instability, and the sudden rise of disk viscosity causes matter to spiral inward (accretion, releasing the outburst energy) and outward, increasing the radius of the disk. During several such complete cycles, the disk's outer edge secularly grows. The next thermal instability (viscosity trigger) pushes the edge out to the 3:1 resonance, where an eccentric (tidal) instability rapidly develops (Whitehurst 1988, Lubow 1991). The latter produces greater dissipation in the now eccentric and larger disk, producing extra light which happens to be modulated with a period slightly longer than $P_{orb}$ (because of precession). The disk then decays on a viscous timescale, and the cycle begins all over again.

The TTI model has been much discussed in the literature, and there are excellent reviews assessing its strengths and weaknesses (Smak 1996, Osaki 1998, Hameury et al. 1998). Here we discuss only the points relevant to the observational record of WZ Sge.

(1) The disk should grow steadily after the viscous trigger ignites on day 13. Yet WZ Sge did not show common superhumps for 13 more days, when the superhumps suddenly turned on and reached high amplitude right away. This must be reckoned surprising. The simulations (Whitehurst 1994, Hirose & Osaki 1990, Lubow 1994, Truss et al. 2000) show a delay, and the delay is greatest at low $q$; but the turn-on should be gradual, whereas the signal appears to reach maximum amplitude in ~1 day.

(2) If the accounting of the OOH in §13 is correct, then the approach to the 3:1 resonance is greatly modified, with the disk reaching the resonance primarily by *contraction*, not expansion. This requires more careful theoretical study, especially to understand the observed very rapid growth of the common superhump.

(3) The TTI model features a constant mass-transfer rate. This is an easy one, because $\dot{M}_2$ is provably enhanced after day 30. The total mass transferred over the era of provable enhancement (essentially the area under Figure 18, converted to energy and mass) is only $\sim 4\times10^{22}$ g, so it is only a small correction to the overall mass budget. However, the enhancement of $\dot{M}_2$ moves the disk into a quite different domain, so this could deeply affect the disk's predicted behavior during decline.

(4) In a simple TTI model, the outburst lasts a long time because the large eccentric disk tidally feeds angular momentum back to the orbit, causing gas to spiral in to the white dwarf and keep the disk bright. Hence the outburst should end when the eccentricity dies. This is contrary to observation, which shows superhumps lasting at least 100 days after the main outburst ends. The TTI model appears to need some revision for this decoupling of superhump and superoutburst (see Hellier 2001).

(5) Finally, it's a curious fact that hot-spot eclipses turned on just as common superhumps rose quickly to prominence, around day 29±2. This is a circumstance which theorists might wish to ponder. It may represent evidence that a sudden rise in $\dot{M}_2$ plays an important role in the rise of superhumps.

## 15. SUMMARY

1. We report light curves of WZ Sge during its 2001 superoutburst. For the first 12 days, a powerful signal at or very near $P_{orb}$ rumbled throughout the light curve. The waveform was double-humped, with a primary minimum at $\varphi$=0.90±0.02. The amplitude declined by 36% each day (hence an e-folding time of 3.2 days). The energy in this signal totaled ~15% of the entire radiant energy of the outburst. The signal's origin remains unknown. However, the theory described above (the two-armed spiral suggested by OM, by the spectroscopy, and by the hydrodynamic calculations) can produce a strong orbital signal at high binary inclination, quench it pretty rapidly, and confine it to dwarf novae of low *q*. These are points of high merit.

2. On day 26 (the 13th day of outburst), another signal rapidly rose in amplitude: the *common superhump*, a hallmark of all short-period dwarf novae in superoutburst. This signal essentially persisted through the 53-day outburst, and at least another 60 days beyond. The period during the main outburst was 0.05721(5) d, lengthening to 0.05738(4) d during the final decline. The exact pattern of period evolution is somewhat hard to specify (see Figures 8 and 17), with complications due to phase shifts possibly arising from absorption in the binary. Probably a transition to “late” superhumps occurred between days 40 and 54.

3. The main outburst lasted 25 days, and was followed by 12 remarkable “echo” outbursts of 1–1.5 mag amplitude. Each rose in ~0.3 d and fell in ~1.2 d. After the 12th echo, the star began a slow decline to quiescence at ~0.02 mag/day.

4. *Mutatis mutandis*, this sequence of events could be described as a faithful reprise of the 1978 outburst. But favorable seasonal timing and coverage by many observing stations in 2001 enabled sensitive measures of the periodic signals, which were found to possess a detailed fine structure. Seventeen noncommensurate signals were found; nearly all were linear combinations of $\omega_o$ (the orbital frequency) and $\Omega$ (the putative frequency of apsidal advance). The detected signals occurred at $n\omega_o-m\Omega$, where $n$=1, 2, ..., 9, and $m$=0, 1, ..., $n$–1. The common superhump began simply, with detections only at $\omega_o-\Omega$, $2\omega_o-2\Omega$, $2\omega_o-\Omega$, and $3\omega_o-\Omega$. During its second week, the superhump attained greater complexity: the strongest components occurred at $n\omega_o-\Omega$ (for $n$=1, 2, ..., 9), with other components at smaller amplitude. During the echo outbursts and final decline, which we associate with ‘late’ superhumps, the structure changed again, favoring components with $n\omega_o-(n-1)\Omega$. The latter may be a consequence of the switch to late superhumps.

5. There was a plausible detection, on several occasions, of a signal at 35.42±0.02 c/d. This could be interpreted as a “negative superhump”, a $2\omega_o+N$ signal indicative of nodal regression at $N$=0.14 c/d. It could alternatively be the first harmonic of an unseen signal at 17.71 c/d. Or it could be something else, which we have not managed to dream up.

6. The data reveal limits on the underlying masses ($M_2<0.08\ M_\odot$ and $M_1>0.8\ M_\odot$), and we adopt $M_1=1.0\pm0.2\ M_\odot$. The value of $M_2$ is less well constrained but is consistent with recent estimates of $q$ (0.057±0.017, Steeghs et al. 2001; 0.045±0.020, Patterson 2001).

7. For $d$=43±8 pc and $M_1=1.0\pm0.2\ M_\odot$, we estimate an accreted mass of $4\times10^{23}$ g in outburst, and a total mass transfer of $10^{24}$ g over the preceding 30 years. Thus the gross energetics do not demand enhanced mass transfer. The main accretion event appears simple enough, just Osaki’s original (1974) theory: the bathtub fills up in quiescence, and empties in outburst.

8. The hot-spot luminosity (or perhaps a lower limit to it) can be estimated from the eclipse depth, and from the height of the orbital wave. These establish that the spot is enhanced over its quiescent luminosity by a factor of ~50 when it becomes first clearly visible around day 30. It then declines with an e-folding time of ~15 days. This requires a greatly enhanced mass-transfer rate from the secondary, and supports the model of Hameury et al. (2000) for the origin of echoes.

9. The sudden appearance of eclipses and orbital humps near day 30 coincides with the rapid growth of common superhumps. This naturally suggests that enhanced $\dot{M}$ may play a major role in superhump growth, a possibility generally overlooked by theorists (with the prominent exception of Whitehurst & King 1991).

10. The observed properties of the eclipse (depth, width, timing) imply an eccentric disk, progradely advancing on the beat period between orbit and superhump. An eccentricity $e \geq 0.3$ is needed to give the variation in eclipse width.

11. Some revisions to the TTI model appear necessary. As previously argued by Smak and Hameury, corrections are needed for the heating of the secondary and consequent enhancement of $\dot{M}_2$. As previously argued by Hellier (2001), the simple TTI model extinguishes superhumps much too early, and it would be nice to have a quantitative understanding of this. The greatest item on the wish list is knowledge of the changing disk radius — which is really needed to test the TTI theory, but not yet well constrained by observation. This is likely to come from intensive observation of eclipsing systems in and near superoutburst (perhaps even data already in hand, somewhere, for WZ Sge).

What a wonderful treat it has been: a mid-summer superoutburst of the world's brightest, nearest, and most celebrated dwarf nova. Eclipsing, and transiting local meridians near midnight. In this paper we have reaped some benefits from this flourish of cosmic philanthropy. Others will certainly follow from intensive campaigns carried out with spectroscopy, and with spaceborne UV and X-ray telescopes. Still others, the most important, will come in the fullness of time, after we have all had the chance to meditate on the great harvest of information. We would like to express our gratitude to: T. Ohshima for his brilliant discovery of WZ Sge on its steep rise to maximum; T. Kato and the VSNET team for their boundless energy in organizing many of the observations from variable-star astronomers around the world, keeping the communications fast and the excitement high; John Cannizzo, Matt Wood, and John Thorstensen for discussions; Jim Kern, Matt Aggleton, Kevin Beaulieu, Dustin Crabtree, Brad Conrad, Marko Moilanen, Harri Hyvonen, Cindy Foote, Jennie McCormick, Fred Velthuis, Tim Hager, Kosmas Gazeas, Alexander Yushchenko, James Hannon, Dan Kaiser, Franco Mallia, and Lasse Jensen for other contributions of data to this enterprise; a wise and anonymous referee for suggestions. This research was supported in part by grants 00–98254 from the NSF and GG–0042 from the Research Corporation.

TABLE 1
LOG OF OBSERVATIONS

| HJD Start (2,452,100+) | Duration (d) | <V> | Observers[a] |
|---|---|---|---|
| 14.4420 | 0.121 | 8.36→8.22 | GM(1), LJ |
| 15.3295 | 0.789 | 8.39 | GM(1), DRS, WA, MWR, PN |
| 16.3140 | 0.812 | 8.67 | GM(1), WA, RR, PN |
| 17.3847 | 0.716 | 8.72 | GM(2), AO, BM, RR, WA, FJ, MWR, SW, PN |
| 18.3086 | 0.593 | 8.94 | GM(1), AO, RN, LMC, PN, AAH |
| 19.3110 | 0.629 | 9.12 | GM(1), RN, REF, LMC, PN, AAH |
| 20.6266 | 0.476 | 9.25 | DS, RR, WA |
| 21.3476 | 0.754 | 9.39 | RN, DK, RR, MWR, DRS, WA, DS |
| 22.3166 | 0.634 | 9.50 | GM(1), MWR, BM, DRS, DS, PN |
| 23.3081 | 0.656 | 9.58 | GM(2), DRS, BM |
| 24.3010 | 0.799 | 9.71 | GM(1), BM, LMC, RR |
| 25.2952 | 0.788 | 9.75 | GM(1), DS, LMC, RR, SW |
| 26.2905 | 0.800 | 9.86 | GM(1), MWR, DS, LMC, BM, RR, SW |
| 27.2902 | 0.596 | 9.90 | GM(1), DS, MWR |
| 28.2898 | 0.678 | 9.97 | GM(1), DS, BM |
| 29.3419 | 0.702 | 10.02 | GM(1), DS, MWR |
| 30.2857 | 0.350 | 10.05 | GM(1) |
| 31.2899 | 0.680 | 10.08 | GM(1), TV, BM |
| 32.2873 | 0.593 | 10.28 | GM(1), TV, MWR, DS |
| 33.2848 | 0.803 | 10.35 | GM(1), TV, MWR, BM, SW, WA, RR |
| 34.2946 | 0.662 | 10.47 | GM(1), BM, DS |
| 35.2989 | 0.548 | 10.58 | GM(1), DS |
| 36.2900 | 0.730 | 10.62 | GM(1), DRS, MWR, WA, BM, DS, AAH |
| 37.4202 | 0.649 | 10.72 | GM(1), MWR, AO, BM, RR, AAH |
| 38.6846 | 0.280 | 11.35→11.53 | BM, LMC, WA, AAH |
| 39.4012 | 0.496 | 11.77→11.99 | TV, LMC, DS, AAH |
| 40.3490 | 0.533 | 12.45→12.61 | AO, GM(1), LMC, EB, AAH |
| 41.3194 | 0.675 | 12.75 | GM(1), AO, OB, BM, SW, WA |
| 42.3454 | 0.634 | 12.63→11.65 | AO, OB, DS, WA |
| 43.3669 | 0.681 | 10.68→11.00 | TV, GM(2), MWR, REF, DH, DRS, RR, WA, SW, AAH |
| 44.3858 | 0.540 | 11.33→11.76 | TV, REF, DH, SW |
| 45.3344 | 0.616 | 12.33→12.05 | TV, DS, EB |
| 46.3297 | 0.697 | 11.23→11.60 | TV, MWR, DS, LMC, SW, WA, JF |
| 47.2915 | 0.652 | 12.05→11.43 | GM(1), TV, DH, DRS, LMC, EB |
| 48.5516 | 0.372 | 11.64→11.82 | DS, BM, EB |
| 49.3216 | 0.511 | 12.33→11.33 | TV, MWR |
| 50.3157 | 0.595 | 11.27→11.68 | TV, MWR, DS, LMC, BM |
| 51.3107 | 0.625 | 12.0→11.3→11.4 | TV, MWR, DS, LMC, BM |
| 52.7006 | 0.161 | 11.92 | LMC, AAH |
| 53.3215 | 0.536 | 11.73→11.23 | TV, MWR, JF, LMC, DRS, BM, DS, OB |
| 54.3181 | 0.548 | 11.78→12.11 | DRS, MWR, JF, JH, LMC |
| 55.5295 | 0.329 | 11.30→11.50 | DRS, MWR, JF, BM, LMC |
| 56.5204 | 0.357 | 12.2→12.2→11.5 | MWR, BM, LMC |
| 57.5548 | 0.319 | 11.23 | EB, DS |
| 58.5155 | 0.391 | 11.21→11.36 | DRS, BM |
| 59.5388 | 0.337 | 11.76→12.23 | DRS, JK(N1), DS |
| 60.5305 | 0.348 | 11.21→11.46 | DRS, JK(N1) |
| 61.5192 | 0.358 | 12.12→11.82 | DRS, JK(N1) |
| 62.5922 | 0.305 | 11.57→11.88 | JK(N1), EB |
| 63.6024 | 0.300 | 12.67 | EB, BM, REF, JK(N1) |
| 64.5183 | 0.379 | 11.33 | DRS, REF, JK(N1), EB, DS |
| 65.5141 | 0.520 | 11.84→12.71 | DRS, EB, WA, DS |
| 66.5040 | 0.423 | 13.21 | DRS, BM, WA |
| 67.5113 | 0.385 | 13.39 | DRS, BM |
| 68.5110 | 0.336 | 13.54 | DRS, JK(N1), DS |
| 69.5087 | 0.444 | 13.77 | DRS, JK(N1), WA, DS |
| 70.2559 | 0.690 | 13.78 | GM(1), AO, DRS, WA |
| 71.5191 | 0.276 | 13.84 | DRS, JF |
| 72.2806 | 0.676 | 13.72 | GM(2), BM, EB, WA |
| 73.6052 | 0.243 | 13.70 | BM, REF, EB |
| 74.2789 | 0.590 | 13.72 | TV, BM, REF |
| 75.6734 | 0.184 | 13.75 | REF |
| 76.2802 | 0.586 | 13.79 | TV, DRS, REF, EB |
| 77.3310 | 0.491 | 13.85 | HH, JF, EB |
| 78.3527 | 0.414 | 13.89 | TV, BM, EB |
| 79.5618 | 0.205 | 13.97 | BM, TK, TC |
| 80.5785 | 0.236 | 14.00 | TK, REF, TC |
| 81.3577 | 0.458 | 13.96 | AO, TV, TK |
| 82.5586 | 0.249 | 13.99 | TK, LMC |
| 83.3476 | 0.462 | 13.94 | AO, TC |
| 84.4917 | 0.312 | 13.85 | DRS, BM, TC |
| 85.5942 | 0.159 | 13.95 | BM |
| 86.2434 | 0.550 | 13.95 | EP, DRS, JF |
| 87.2882 | 0.502 | 13.94 | EP, TV, DRS, BM, TC |
| 88.5857 | 0.132 | 13.93 | BM, TC |
| 89.5021 | 0.294 | 13.92 | DRS, TK |
| 90.4800 | 0.297 | 13.94 | DRS, TK, TC |
| 91.2735 | 0.336 | 13.91 | TV, DRS |
| 92.2678 | 0.471 | 13.96 | TV, DRS, LMC |
| 93.3245 | 0.469 | 13.99 | TV, JK(S), EB, REF |
| 94.2721 | 0.280 | 14.04 | TV, JK(S), TK, LMC |
| 95.2815 | 0.511 | 14.07 | TV, JK(S), TK, EB, REF, LMC |
| 96.3035 | 0.486 | 14.15 | TV, TK, EB, JF |
| 97.2643 | 0.477 | 14.20 | AO, REF, TC, AAH |
| 98.4762 | 0.315 | 14.24 | DRS, EB, NM |
| 99.4795 | 0.290 | 14.23 | JK(S), JF, NM |
| 101.4640 | 0.121 | 14.28 | DRS |
| 102.4861 | 0.273 | 14.35 | DRS, EB |
| 103.4784 | 0.293 | 14.31 | DRS, EB |
| 104.5314 | 0.074 | 14.39 | NM |
| 105.4835 | 0.100 | 14.41 | JK(S), DRS |
| 106.4871 | 0.106 | 14.43 | JK(S) |
| 107.5689 | 0.185 | 14.45 | EB |
| 108.1976 | 0.545 | 14.43 | EP, EB |
| 109.2166 | 0.526 | 14.49 | EP, EB, TC |
| 110.2347 | 0.093 | 14.38 | EP |
| 111.5093 | 0.223 | 14.40 | TC |
| 112.4552 | 0.262 | 14.39 | DRS, TC |
| 145.5745 | 0.056 | 14.60 | JK(N1) |
| 146.5654 | 0.058 | 14.52 | JK(N1) |
| 147.5552 | 0.082 | 14.54 | JK(N1) |
| 174.5782 | 0.007 | 14.62 | JK(N2) |

NOTES. — Two or more $V$ estimates supplied for nights with strong secular trends. See text for caveats about accuracy. Run duration is calculated for each night from beginning to end. Data typically spans ~80% of this interval.

[a]Observer: AAH = USNO 1.0 m, A. Henden; AO = CBA–Finland 41 cm, A. Oksanen, M. Moilanen, & H. Hyvonen; BM = KUC 30 cm, B. Martin; DH = CBA–Hamburg 41 cm, D. Husar; DRS = CBA–East 66 cm, D. Skillman; DS = CBA–Indiana 30 cm, D. Starkey; EB = CBA–Colorado 25 cm, E. Beshore; FJ = CBA–Pakuranga 35 cm; J. McCormick & F. Velthuis; GM(1) = CBA–Italy 28 cm, G. Masi; GM(2) = CAO 80 cm, G. Masi; JF = CBA–Utah 50 cm, J, Foote & C. Foote; JK(N1) = MDM 1.3m, J. Kemp; JK(N2) = MDM 2.4m, J. Kemp; JK(S) = CTIO 0.9m, J. Kemp; LMC = CBA–Concord 44 cm, L. Cook; MWR = RIT 25 cm, M. Richmond, S. Davis, T. Davis, J. Kern, M. Aggleton, K. Beaulieu, D. Crabtree, & B. Conrad; OB = CBA–Huntley 28 cm, O. Brettman; PGN = CBA–Greece 41 cm, P. Niarchos, K, Gazeas, & A. Yushchenko; REF = CBA–Braeside 41 cm, R. Fried; RN = NCO 41 cm, R. Novak, et al.; RR = CBA–Nelson 35 cm, R. Rea; SW = CBA–Waiharara 30 cm, S. Walker; TC = ATU 41 cm, T. Campbell & J. Robertson.TH = CBA–New Milford 25 cm, T. Hager; TK = CBA–New Mexico 28 cm, T. Krajci; TV = CBA–Belgium 35 cm, T. Vanmunster; WA = CBA–Blenheim 30 cm, W. Allen.

TABLE 2
OOH AND SUPERHUMP MAXIMA

| (HJD 2,452,100+) | | | | | | | |
|---|---|---|---|---|---|---|---|
| 14.484 | 17.785 | 21.354 | 23.962 | 26.678 | 29.446 | 32.592 | 35.334 |
| 14.540 | 17.840 | 21.410 | 24.357 | 26.736 | 29.560 | 32.650 | 35.388 |
| 15.350 | 17.897 | 21.470 | 24.413 | 26.792 | 29.673 | 32.709 | 35.445 |
| 15.405 | 17.953 | 21.523 | 24.470 | 26.853 | 29.731 | 32.762 | 35.504 |
| 15.4625 | 18.009 | 21.637 | 24.528 | 26.907 | 29.787 | 32.821 | 35.677 |
| 15.5185 | 18.068 | 21.692 | 24.694 | 26.965 | 29.958 | 33.333 | 35.790 |
| 15.573 | 18.357 | 21.804 | 24.752 | 27.021 | 30.301 | 33.389 | 36.308 |
| 15.6335 | 18.408 | 21.980 | 24.812 | 27.079 | 30.358 | 33.447 | 36.364 |
| 15.689 | 18.463 | 22.034 | 24.867 | 27.320 | 30.414 | 33.503 | 36.420 |
| 15.7455 | 18.523 | 22.088 | 24.925 | 27.382 | 30.471 | 33.560 | 36.478 |
| 15.801 | 18.745 | 22.322 | 24.979 | 27.438 | 30.529 | 33.616 | 37.467 |
| 16.084 | 18.803 | 22.374 | 25.036 | 27.496 | 30.586 | 33.731 | 37.523 |
| 16.372 | 18.861 | 22.429 | 25.321 | 27.553 | 31.327 | 33.787 | 37.637 |
| 16.4255 | 19.314 | 22.543 | 25.379 | 27.612 | 31.383 | 33.844 | 37.693 |
| 16.480 | 19.370 | 22.599 | 25.436 | 27.671 | 31.440 | 33.902 | 37.751 |
| 16.538 | 19.428 | 22.713 | 25.719 | 27.727 | 31.497 | 33.958 | 37.809 |
| 16.937 | 19.4815 | 22.827 | 25.884 | 27.787 | 31.554 | 34.015 | 37.865 |
| 16.990 | 19.710 | 22.881 | 25.945 | 27.843 | 31.725 | 34.074 | 37.923 |
| 17.048 | 19.824 | 22.940 | 26.001 | 28.299 | 31.783 | 34.365 | 37.980 |
| 17.105 | 19.879 | 23.335 | 26.060 | 28.356 | 31.839 | 34.419 | 38.036 |
| 17.4445 | 20.735 | 23.393 | 26.338 | 28.413 | 31.896 | 34.477 | |
| 17.502 | 20.790 | 23.507 | 26.396 | 28.470 | 31.954 | 34.534 | |
| 17.557 | 20.848 | 23.623 | 26.451 | 28.527 | 32.367 | 34.589 | |
| 17.613 | 20.958 | 23.676 | 26.509 | 28.698 | 32.423 | 34.705 | |
| 17.672 | 21.016 | 23.737 | 26.566 | 28.925 | 32.480 | 34.761 | |
| 17.728 | 21.073 | 23.850 | 26.621 | 29.390 | 32.537 | 34.932 | |

TABLE 3
PERIODS AND AMPLITUDES OF COMMON SUPERHUMPS

| Day | Period (d) | Amplitude (counts) |
|---|---|---|
| 27– 33 | 0.05719(4) | 41 |
| 31– 37 | 0.05727(6) | 18 |
| 43– 54 | 0.05745(4) | 2.2 |
| 53– 65 | 0.05729(4) | 3.5 |
| 68– 79 | 0.05731(4) | 2.1 |
| 80– 91 | 0.05736(4) | 1.2 |
| 89– 99 | 0.05739(4) | 0.9 |
| 99–109 | 0.05738(4) | 0.9 |

TABLE 4
TIMES OF DEEPEST ECLIPSE
IN THE 5-DAY CYCLE

| (Day number) | | | |
|---|---|---|---|
| 32.3 | 58.7 | 80.7 | 99.5 |
| 37.9 | 64.7 | 85.8 | 104.3 |
| 42.5 | 70.0 | 90.6 | 108.5 |
| 48.0 | 75.4 | 95.2 | |

NOTE. — Estimated error = ±0.4 days.

## FIGURE CAPTIONS

FIGURE 1. — *Upper frame*, eruption light curve of WZ Sge in 2001, showing a rapid rise to $V$=8.2 and subsequent decay at ~0.10 mag/day, punctuated by an episode of 12 "echo outbursts". *Lower frame*, expanded view of the echo episode. A freehand curve has been added to improve visibility.

FIGURE 2. — *Upper frame*, light curve during day 17, dominated by the outburst orbital hump. The light curve was prewhitened by removing the mean and linear trend from the original data. *Lower frame*, orbital light curves for the first 8 days, converted to intensity units with the mean removed. Each day is labeled with the day number. The signal amplitude falls by ~36% each day.

FIGURE 3. — *Upper frame*, power spectrum of day 14–22, with significant features labeled with their frequencies in c/day (all ±0.012 c/day). These appear to be harmonics of a signal at 17.649±0.006 c/day, consistent with the known $\omega_{orb}$. The fundamental of the outburst orbital hump (OOH) rises off-scale to a power of 1130. *Lower frames*, synchronous summations at $P_{orb}$, early and late in the 12-day tenure of the OOH. There are some differences in waveform, but the primary minimum stays at orbital phase 0.90±0.02. The units of power are arbitrary, but proportional to amplitude squared.

FIGURE 4. — Average power spectrum of days 14–18, 18–21, and 21–24, after separate removals of the strong OOH. Possibly significant features are labeled with their frequency in c/day (±0.04).

FIGURE 5. — Orbital light curves during days 26–37. Each frame is averaged over 4–8 binary orbits, and is labeled with the date of mid-observation. Each orbit is resolved into 100 phase bins, with no smoothing. Average magnitudes are given in Table 1.

FIGURE 6. — *Upper frame*, light curve of day 27, dominated by common superhumps. A few very narrow dips in the light curve are also seen, occurring at a fixed binary phase (0.70±0.01). *Middle frames*, power spectrum of the light curve in day 27–33, with significant detections labeled by their frequencies in c/day (all ±0.02 c/day). The signal at 17.49 c/d rises off-scale to a power of 1030. *Lower frames*, synchronous summations at the superhump and orbital periods, day 27–33.

FIGURE 7. — The common superhump, day 31–37. *Upper and middle frames*, the power spectrum, which has been "cleaned" for the strongest signals only (17.46 and 34.93 c/day). Strong signals are flagged by their frequencies in c/day (±0.02). Two are marginal (70.06 and 70.68 c/day). Of the seventeen remaining, the strongest ones are either simple integer multiples of $\omega_o$ and $\omega_o$–$\Omega$, or signals at $n\omega_o$–$\Omega$. *Lower frames*, synchronous summations at the superhump and orbital period; the bright-spot eclipse, a familiar feature in the quiescent orbital light curve, has returned.

FIGURE 8. — O–C diagram of maxima during days 14–37, with respect to the quiescent eclipse

ephemeris. A sharp transition to a different period on day 26 is evident.

FIGURE 9. — Mean (nightly) orbital curves obtained during the dip event, days 38–42. Each frame is tagged with its day number and estimated mean *V* magnitude. The complexity arises from the simultaneous presence, at comparable amplitude, of orbital and superhump signals. The eclipse is seen throughout.

FIGURE 10. — Two consecutive nightly light curves during the echo outbursts.

FIGURE 11. — *Upper and middle rows*, cleaned power spectra of day 43–54 and 53–65, with frequency errors of 0.010 c/day. See text for frequency assignments. But the extreme upper right frame is uncleaned, because the $4\omega_o$ and $4\omega_o$–$4\Omega$ components are spaced by an interval too close to 1.00 c/day to enable separation. *Lowest row*, cleaned power spectrum of day 43–65, with frequency errors of 0.006 c/day. The 17.44 c/day detections in the middle and lowest row rise offscale to a power of 550 and 440, respectively.

FIGURE 12. — Nightly orbital light curves during final decline, with dates of mid-observation identified in each frame. Most frames have 100 phase bins per orbit, with all bins independent (no smoothing). The light curves tend to repeat with a cycle of ~5 days.

FIGURE 13. — See caption of Figure 12.

FIGURE 14. — See caption of Figure 12, except that 50 bins are most commonly used here. On day 100, we let sleep the dogs of war.

FIGURE 15. — *Upper row*, portions of the cleaned power spectrum for days 68–79. There was no significant signal near $3\omega_o$. *Middle and lower rows*, same for days 80–91 and 89–99. Significant peaks or candidates are labeled with their frequencies in c/d (±0.012). Listed at right are the days covered.

FIGURE 16. — Folded light curves on the orbital and superhump periods. Day numbers are attached to the orbital light curves (but apply to both). Zero superhump phase is defined by maximum light, which occurred at HJD 68.5184, 80.6116, and 89.5032.

FIGURE 17. — *Upper frame*, superhump periods during various segments of the outburst, deduced from power spectra. *Lower frame*, O–C diagram of superhump maxima, relative to the test ephemeris HJD 26.6248+0.05727*E*. The time coordinate in the two frames is the same. Inset tick marks show the times of echo outbursts, which have no discernible effect on superhumps.

FIGURE 18. — The changing eclipse depth (measured at deepest eclipse in the 5-day cycle) and orbital wave height (averaged over one or two 5-day cycles), in intensity units. The echo era is shown by the inset tick marks. The eclipse depth is ~0.4 at quiescence, and should be ~4.0 for a mass-transfer rate of $10^{16}$ g/s. These signatures of a brilliant hot spot appear to turn on near day 27.

FIGURE 19. — Constraints on distance modulus and white dwarf mass. The horizontal lines are the parallax constraint (Thorstensen 2001), and the $M_1$>0.77 $M_\odot$ constraint comes from spectroscopy in eruption (Steeghs et al. 2001). The diagonal curve expresses the mass–$M_V$ relation for a 15000 K white dwarf emitting the observed UV/optical flux ascribed to the white dwarf in quiescence. The star may be anywhere on or below this line, since the effective radiating area may be less than a normal white-dwarf hemisphere. Thus the black region satisfies all constraints. At bottom are recent estimates of $M_1$ from photometry and spectroscopy in quiescence.

FIGURE 20. — Estimates of the precession period, based on the running mean of three consecutive "deepest eclipse" timings (from Table 4)

FIGURE 21. — Measured depth, width, and mid-eclipse time of the hot-spot eclipse, for common superhumps (days 31–39, left panels) and late superhumps (days 68–92, right panels). Open circles show upper limits. Zero precession phase is taken to be the time when superhump maximum and orbital eclipse coincide. This produces the conventions $\Phi = (t - 34.135)/5.7$ and $(t - 71.322)/4.95$ for common and late superhumps respectively. The sinusoids are fits to amplitude and phase, with fixed period. See text for discussion of these cycles.

## APPENDIX A
### Hump Zoology in Cataclysmic Variables

| Type | Meaning | Example Stars | Alleged Origin |
|---|---|---|---|
| 1. Orbital Hump | Signal at the orbital frequency $\omega_o$ in quiescence. | U Gem<br>WZ Sge | Presentation effect of hot spot (stream–disk impact region). |
| 2. Outburst Orbital Hump | Signal at or very near $\omega_o$ in outburst. Quite rare; appears to be transiently present in a few SU UMa stars (possibly restricted to the WZ Sge class) in the earliest stage of outburst. Sometimes also called early, immature, and orbital superhumps; but we prefer to restrict "superhump" to cases where the frequency is distinct from $\omega_o$. | WZ Sge<br>AL Com | Unknown. |
| 3. Common Superhump | Signal at $\omega_o-\Omega$, shown by all SU UMa stars in outburst; decays roughly on a timescale of 1–3 weeks. Often of very large amplitude (0.4 mag), and thus a major element in the outburst energy budget. So universal and so extensively studied that "superhump", *sans* qualifier, often implies a common superhump. | SU UMa<br>VW Hyi<br>54 others | Periodic tidal disturbance of the disk by the orbiting secondary (thus requiring a slow apsidal advance to match the frequency shift to $\omega_o-\Omega$). |
| 4. Late Superhump | Signal at $\omega_o-\Omega$, sometimes following (3) and basically defined by a sudden phase shift in (3) of ~0.5 cycles, with little or no change in period. | OY Car<br>VW Hyi | Not securely known, but definitely similar to (3) — features apsidal advance of an "elliptical" disk. |
| 5. Positive Superhump | A general term for any signal with $P$ slightly exceeding $P_o$ (a small positive increment in period) and hence $\omega=\omega_o-\Omega$. Includes all common superhumps. | 77 CVs | Just an observational term. |
| 6. Apsidal Superhump | Alternate to (5), if you subscribe to the theory that (5) arises from apsidal advance ("precession") of the disk, in which case $\Omega$ is the precession frequency. Includes all common superhumps, if you buy that theory. | 77 CVs, probably | Probably same as (3) or (4). |
| 7. Negative Superhump | A general term for any signal with $P$ slightly less than $P_o$ (a small negative increment in period) and hence $\omega=\omega_o+N$. | V503 Cyg<br>TV Col<br>V603 Aql | Just an observational term. |
| 8. Nodal Superhump | Alternate to (7), if you subscribe to the theory that (7) arises from nodal precession (wobble) of the disk, in which case $N$ is the precession frequency. | same, probably | Not securely known. |
| 9. Permanent Superhump | Any positive or negative superhump which is long-lived (months or longer) and not associated with eruption. | AM CVn<br>V603 Aql<br>BK Lyn | Probably same as (3), (4), and (8). |
| 10. Quiescent Superhump | Extremely rare, and not a term in general use. A superhump in states of very low luminosity, with no connection yet established to the other superhump types. | AL Com<br>CP Eri | Unknown. |
| 11. Superhumper | A star which engages in superhumps. | | |

Notes. —

(a) "Outburst" here means *superoutburst*. Happily, we still know of no related periodic signals characteristic of normal outburst.

(b) It may well be true that all apsidal precession is prograde (giving a positive superhump) and all nodal precession is retrograde (giving a negative superhump). The limited data available now are consistent with this. If counterexamples are found, these definitions would be affected somewhat.

(c) Superhumps can be characterized by 3 fundamental frequencies ($\omega_o$, $\Omega$, $N$), and the dominant signal is nearly always $\omega_o-\Omega$ or $\omega_o+N$. But studies of high sensitivity and frequency resolution often reveal components with $\omega=n\omega_o-m\Omega$ (where $n$=any small integer and $m$=1, 2, ..., $n$) or $n\omega_o+N$ (same terminology). We consider these as "fine structure" and thus outside this classification effort.

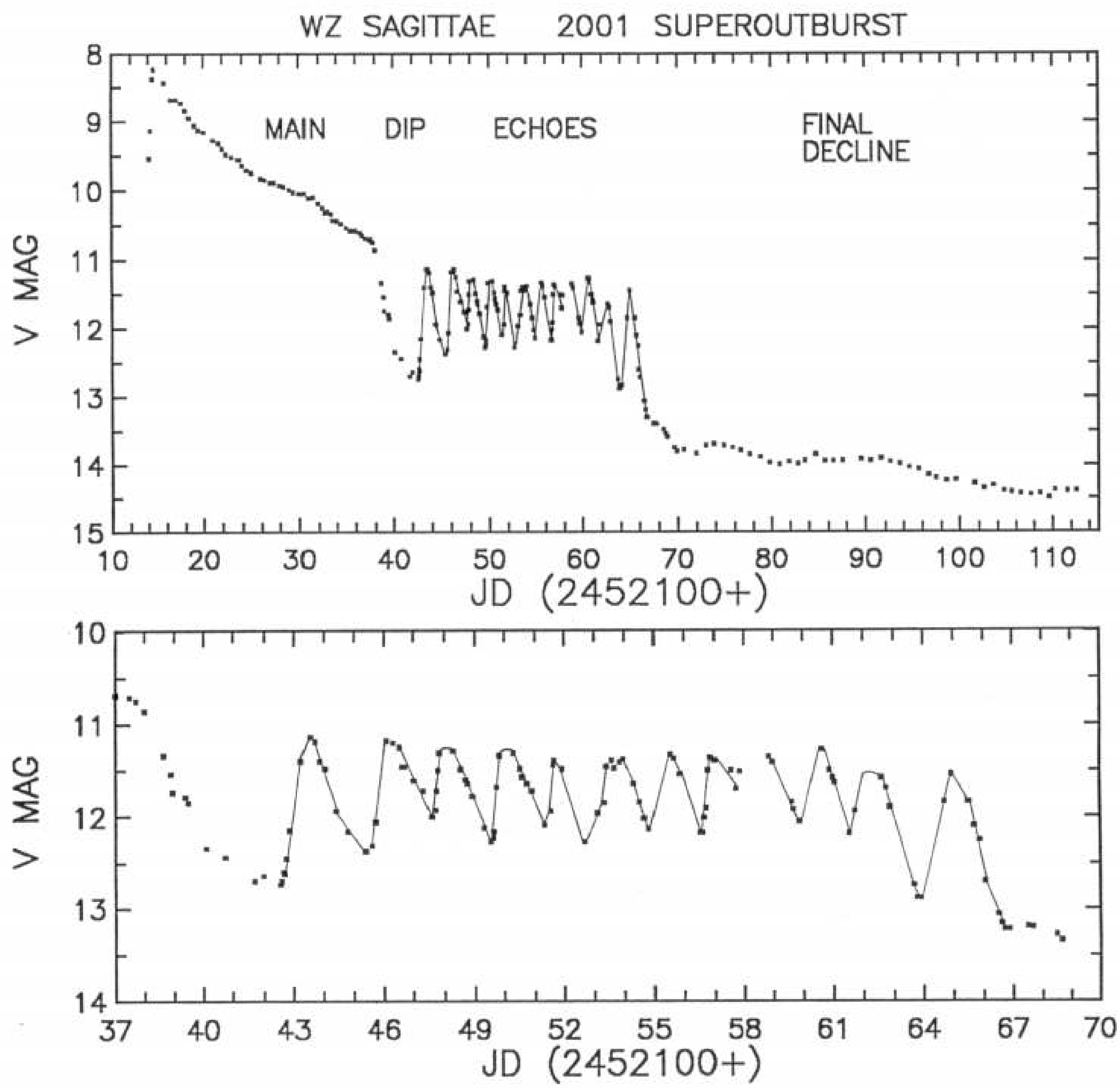

Fig 1

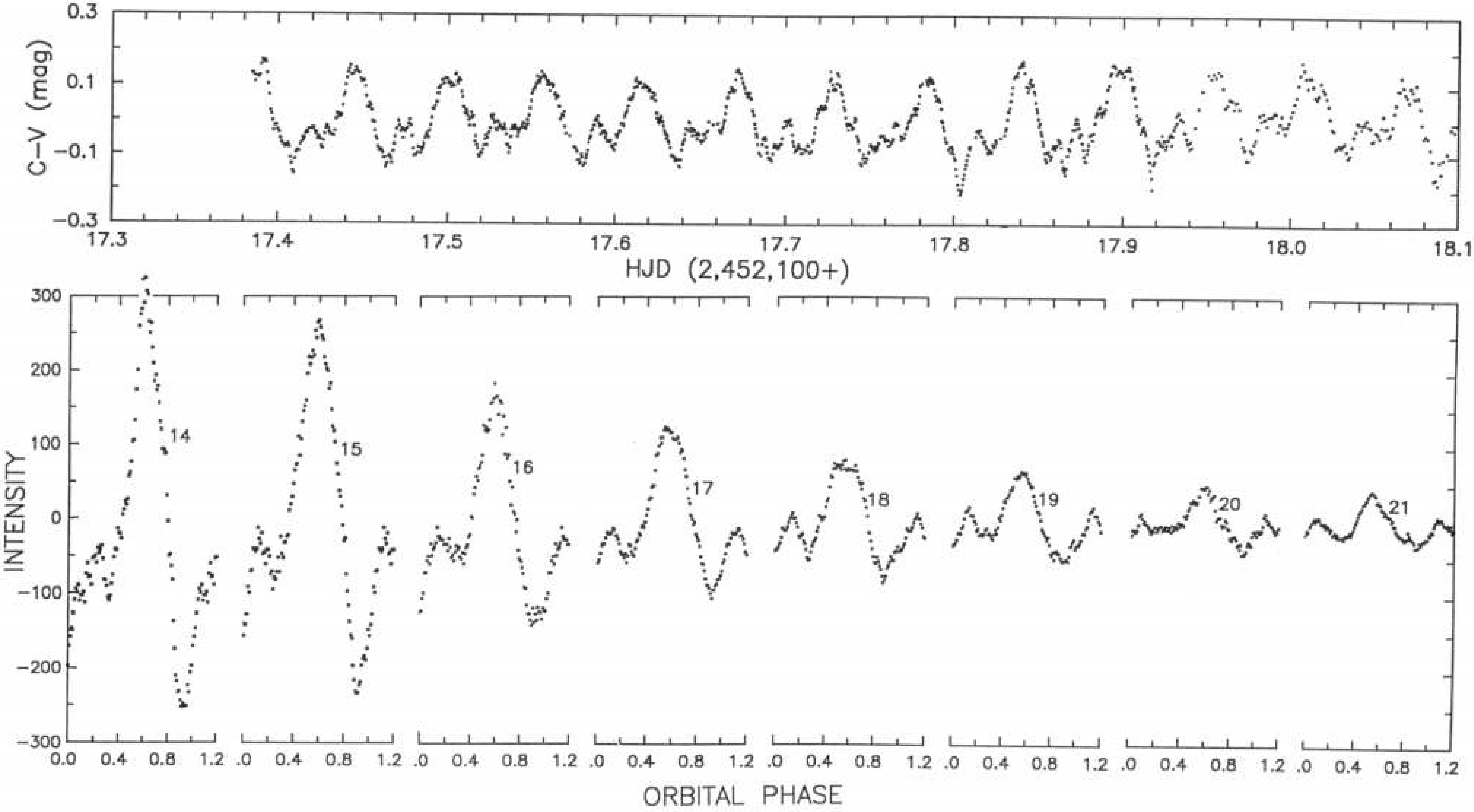

Fig 2

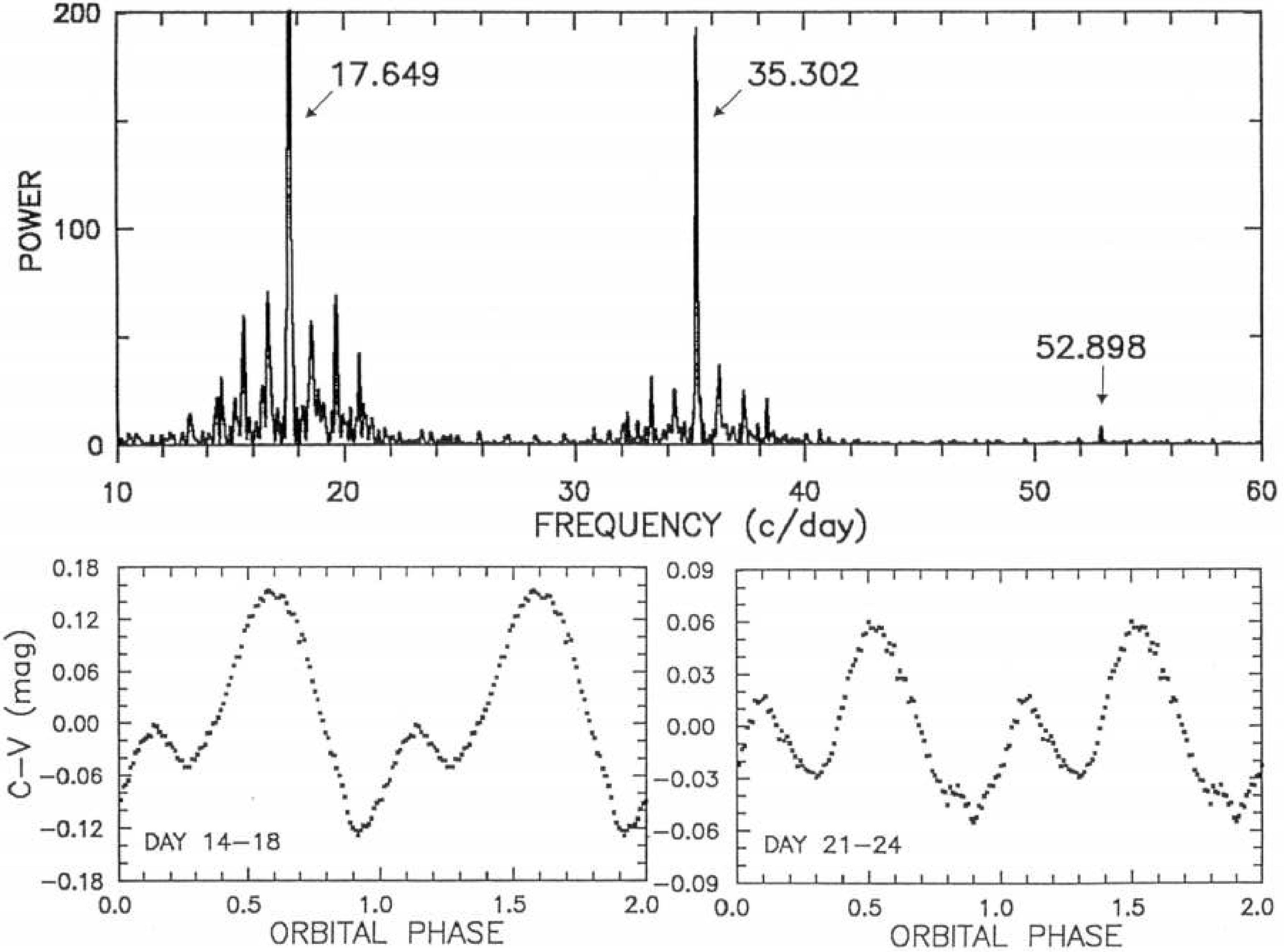

Fig 3

Fig 4

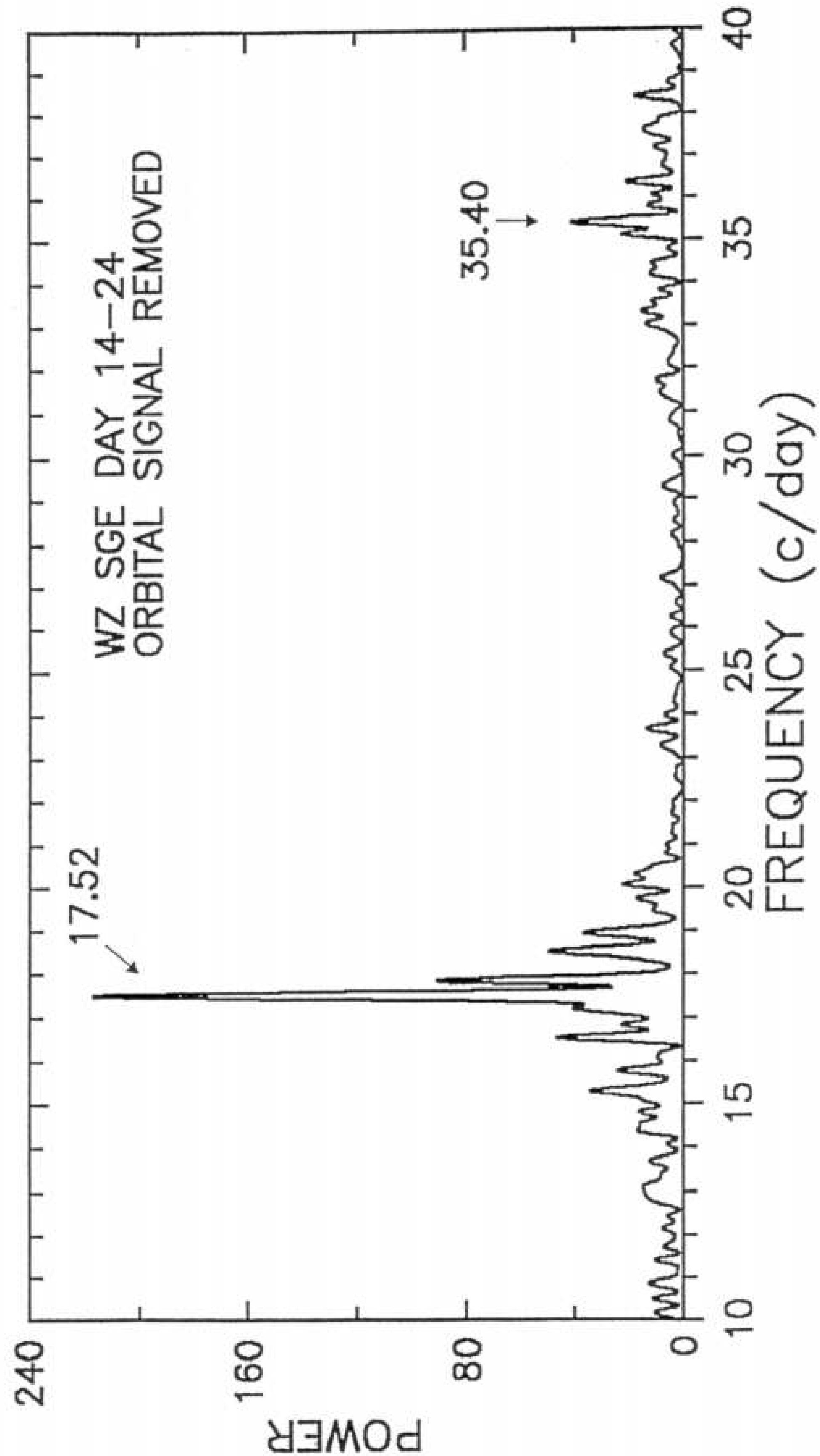

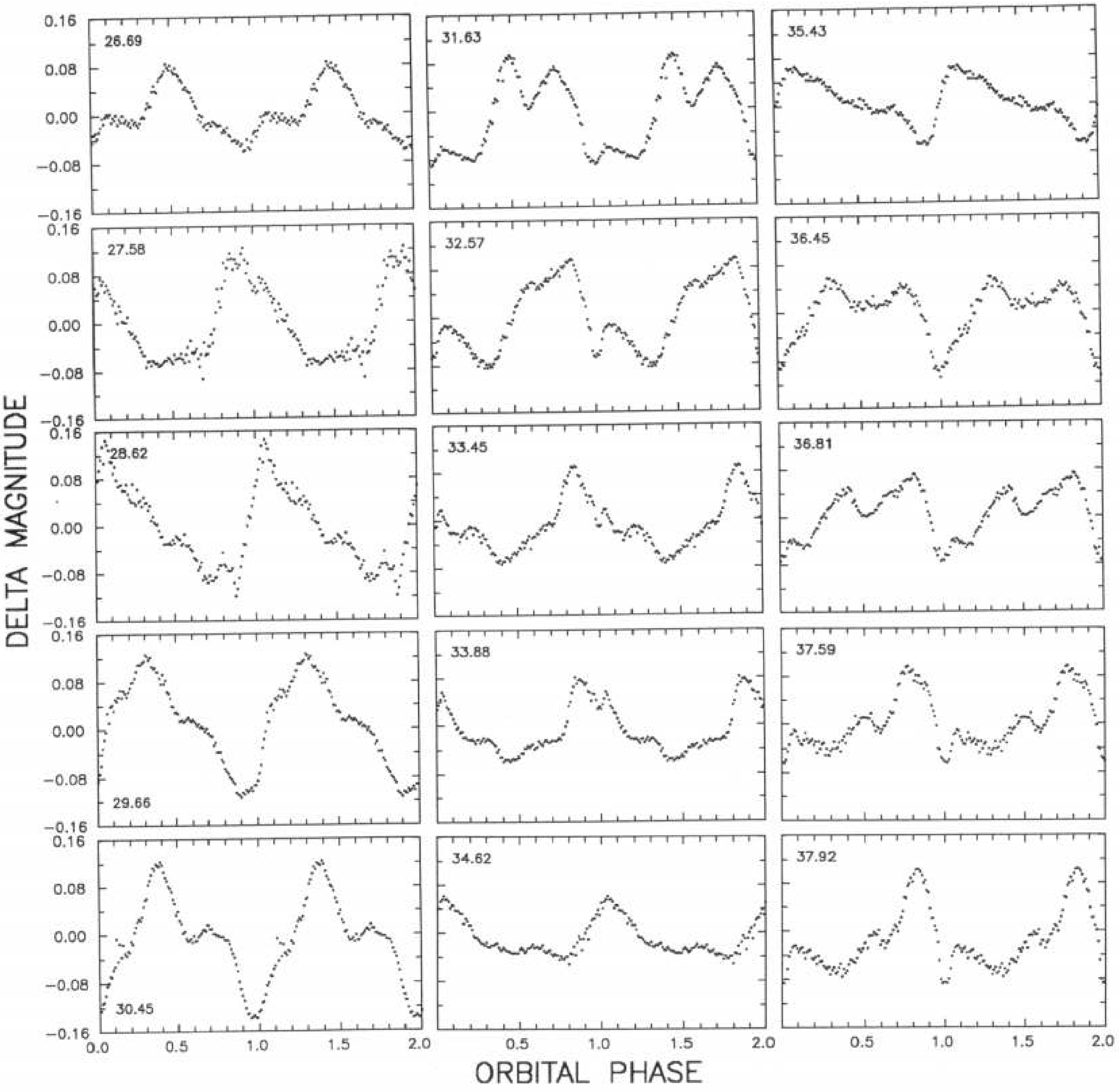

Fig 5

Fig 6

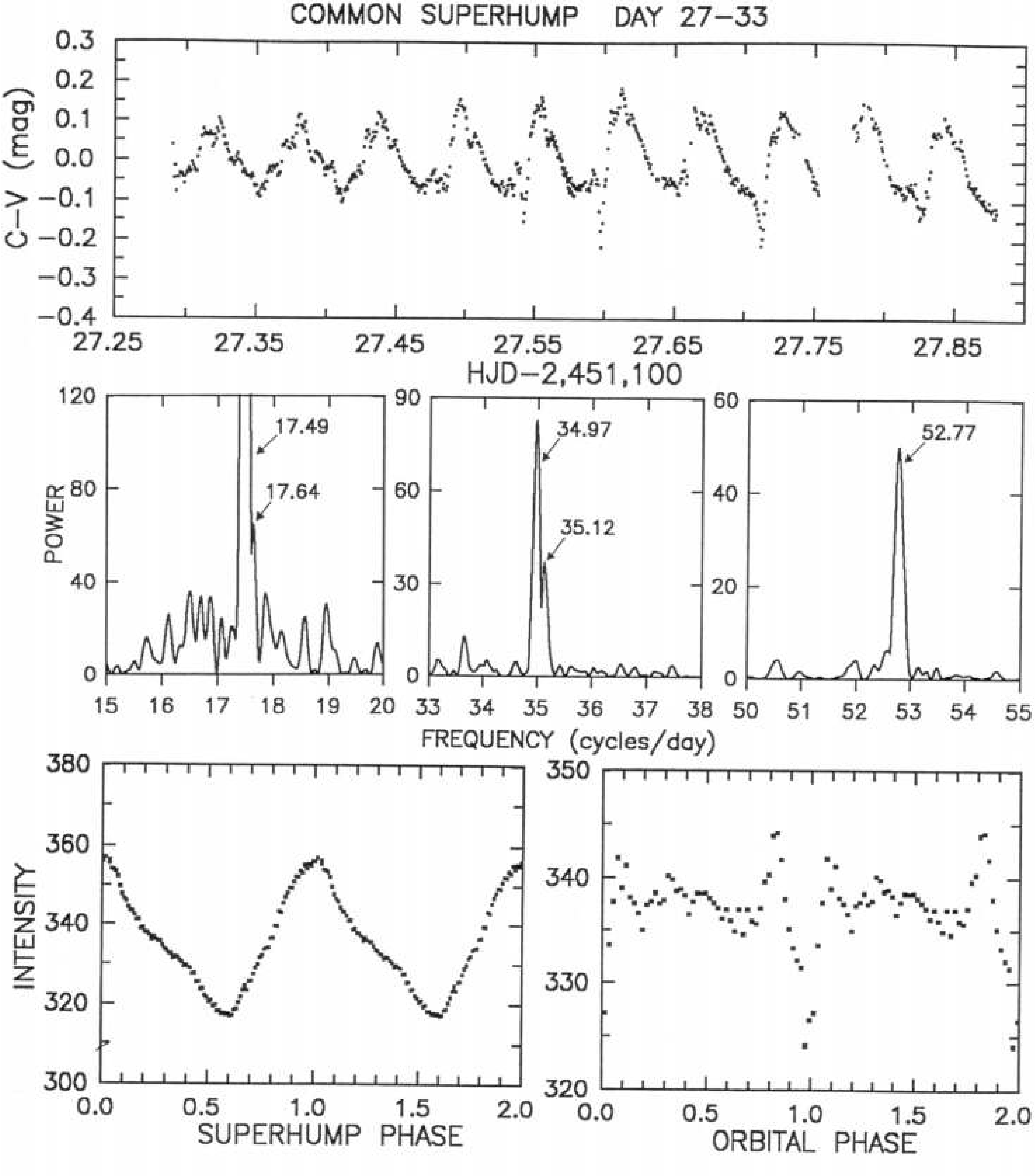

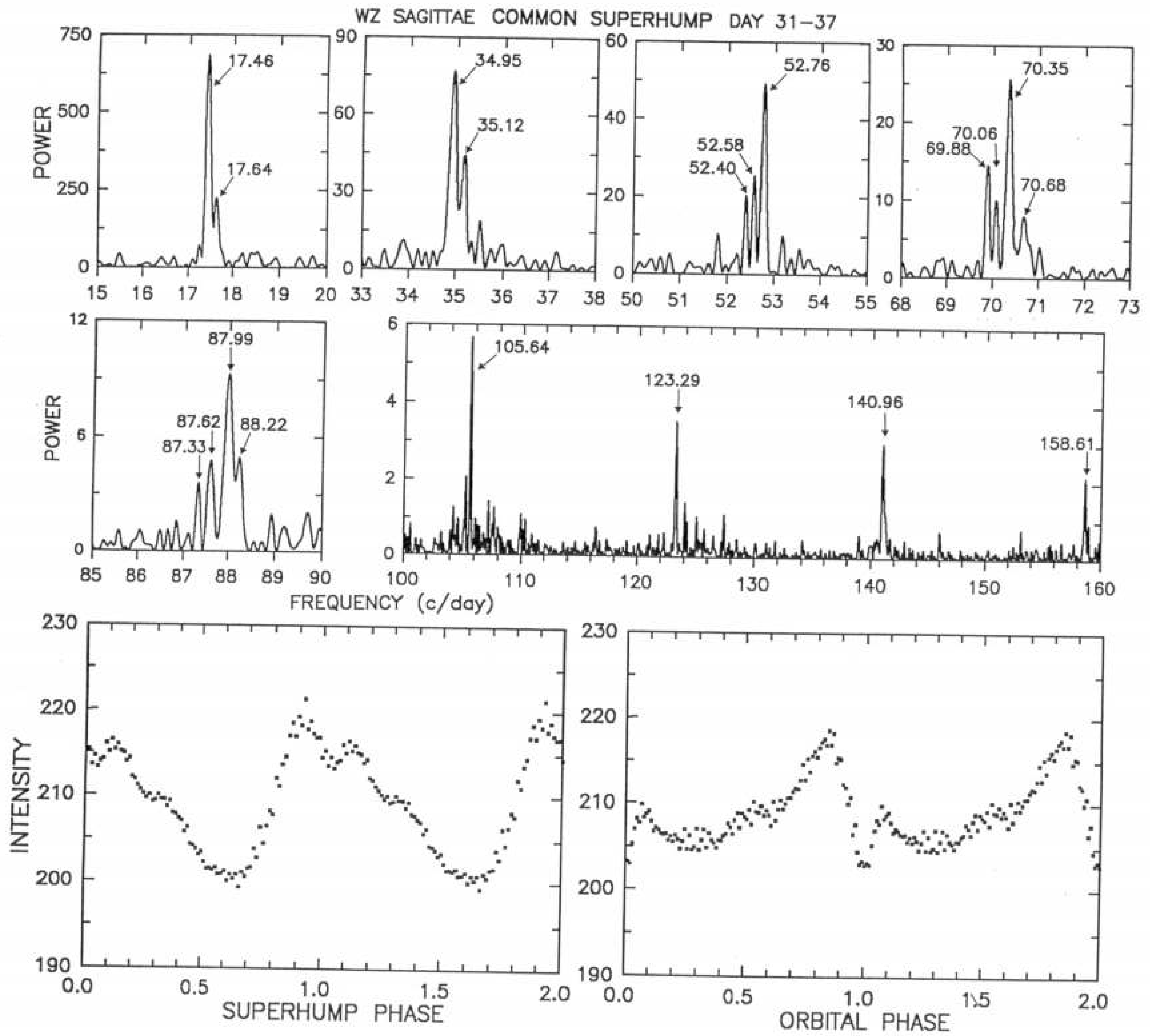

Fig 7

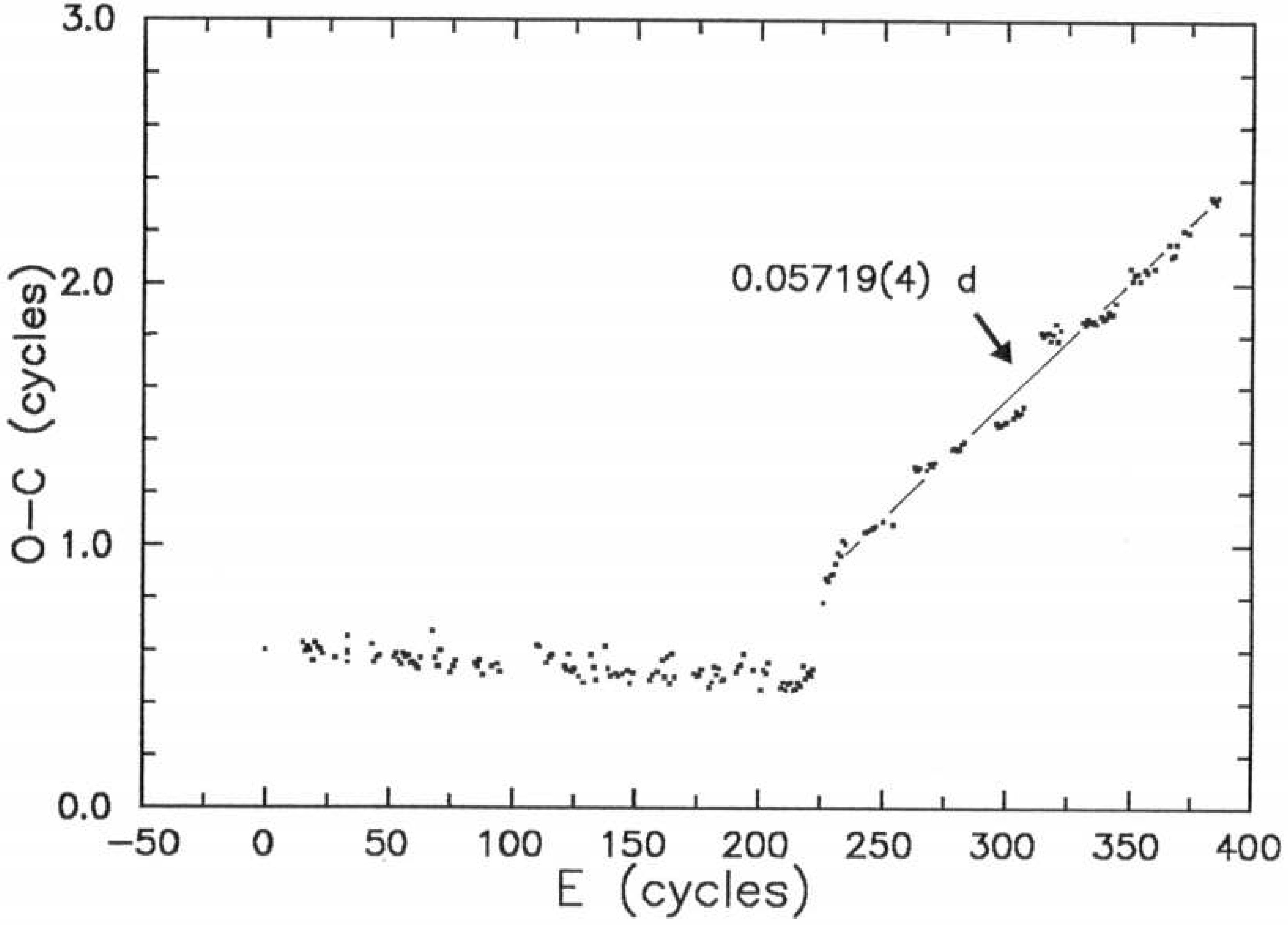

Fig 8

Fig 9

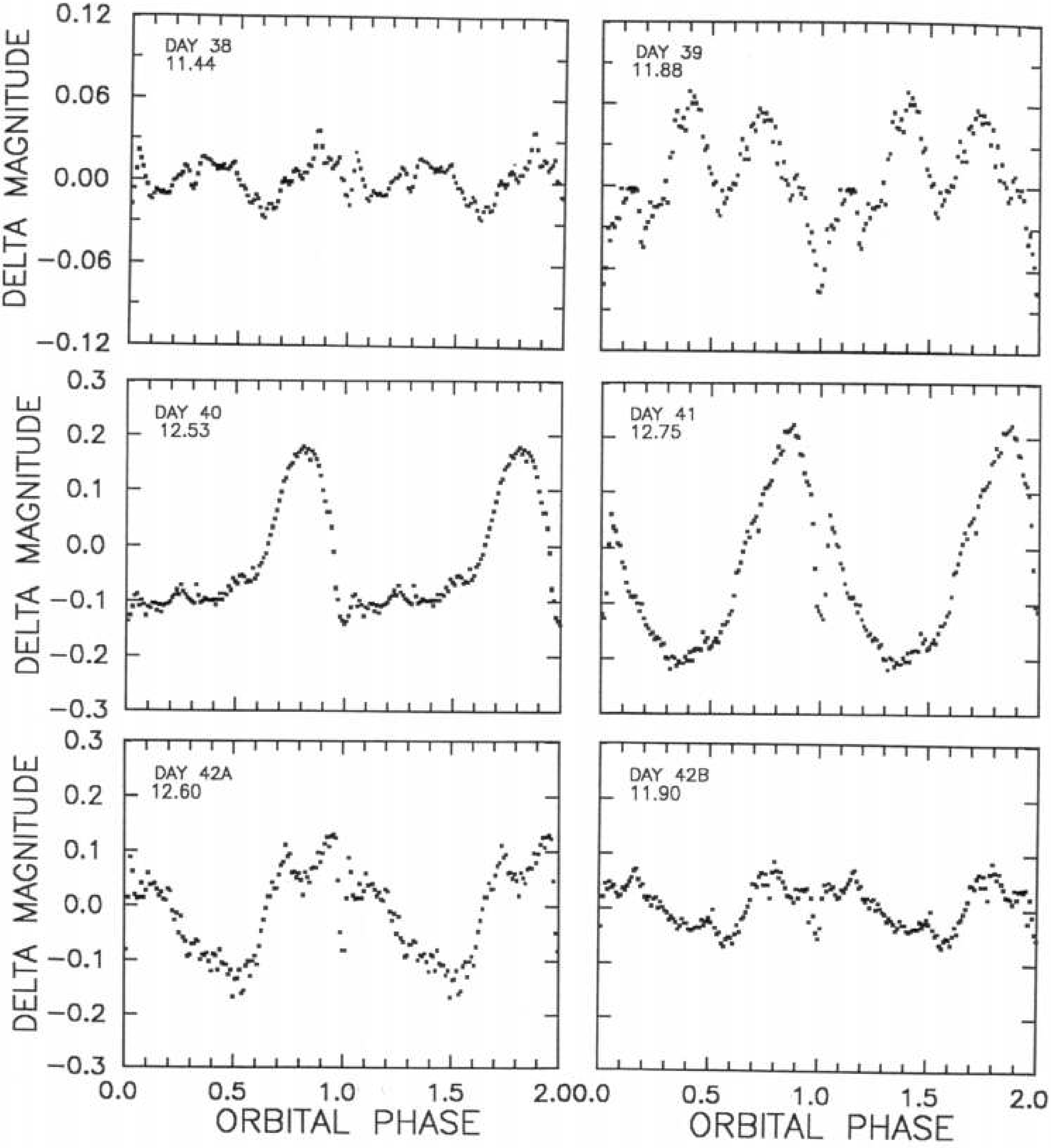

Fig 10

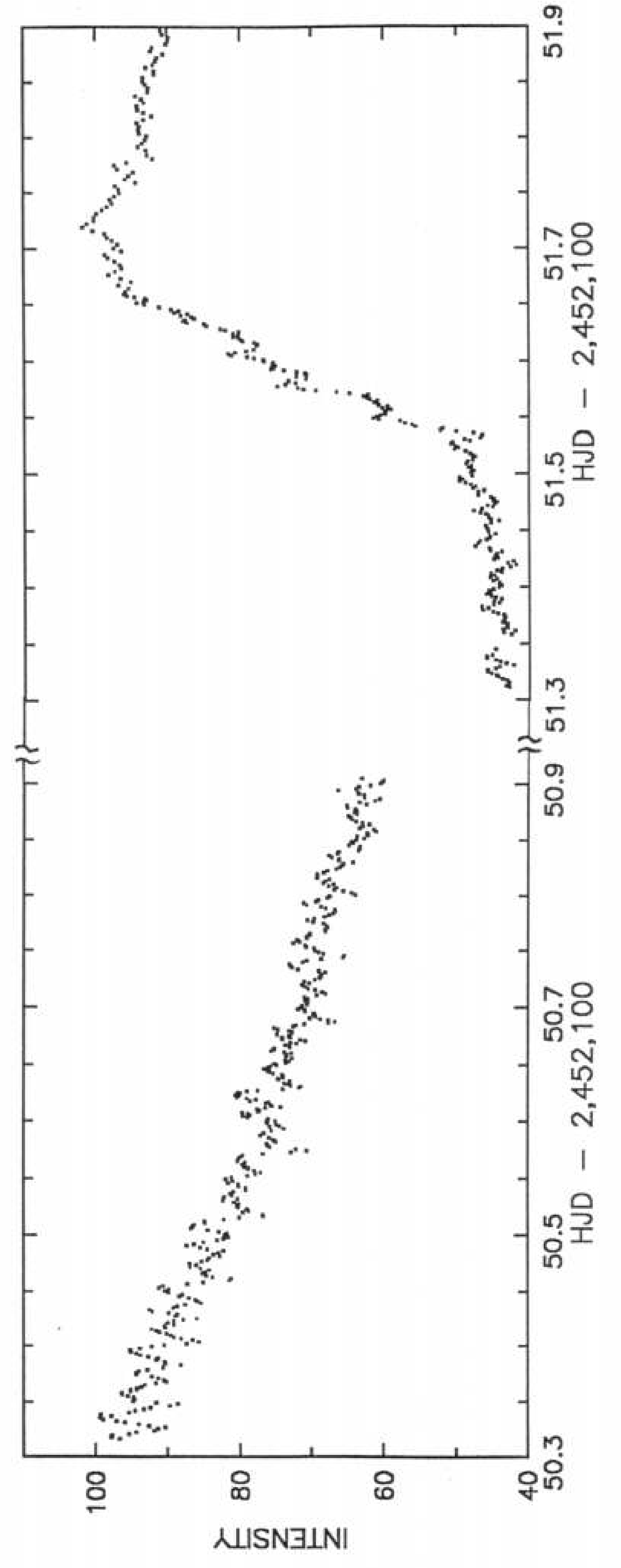

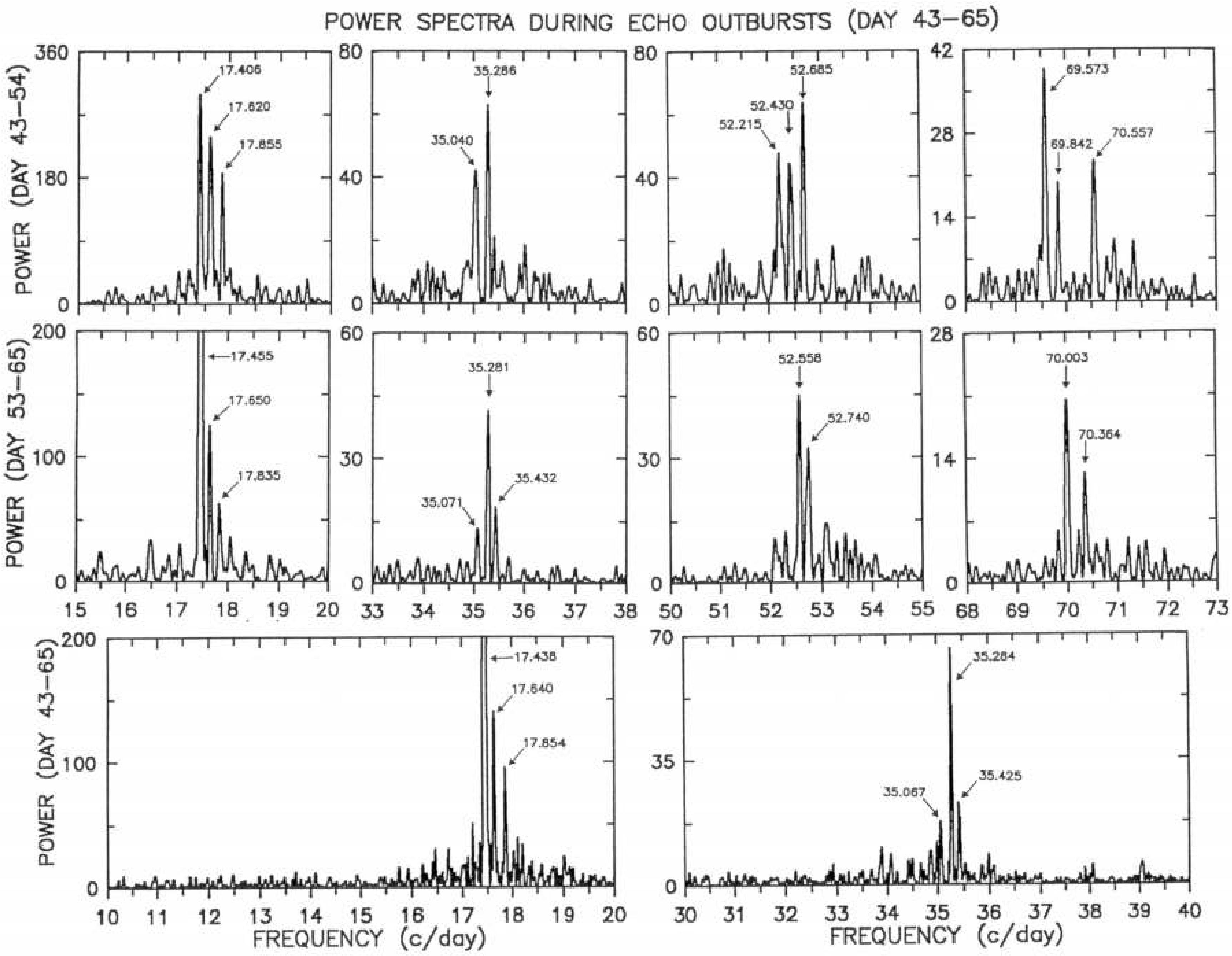

Fig 11

Fig 12

WZ SAGITTAE NIGHTLY ORBITAL LIGHT CURVES DURING FINAL DECLINE

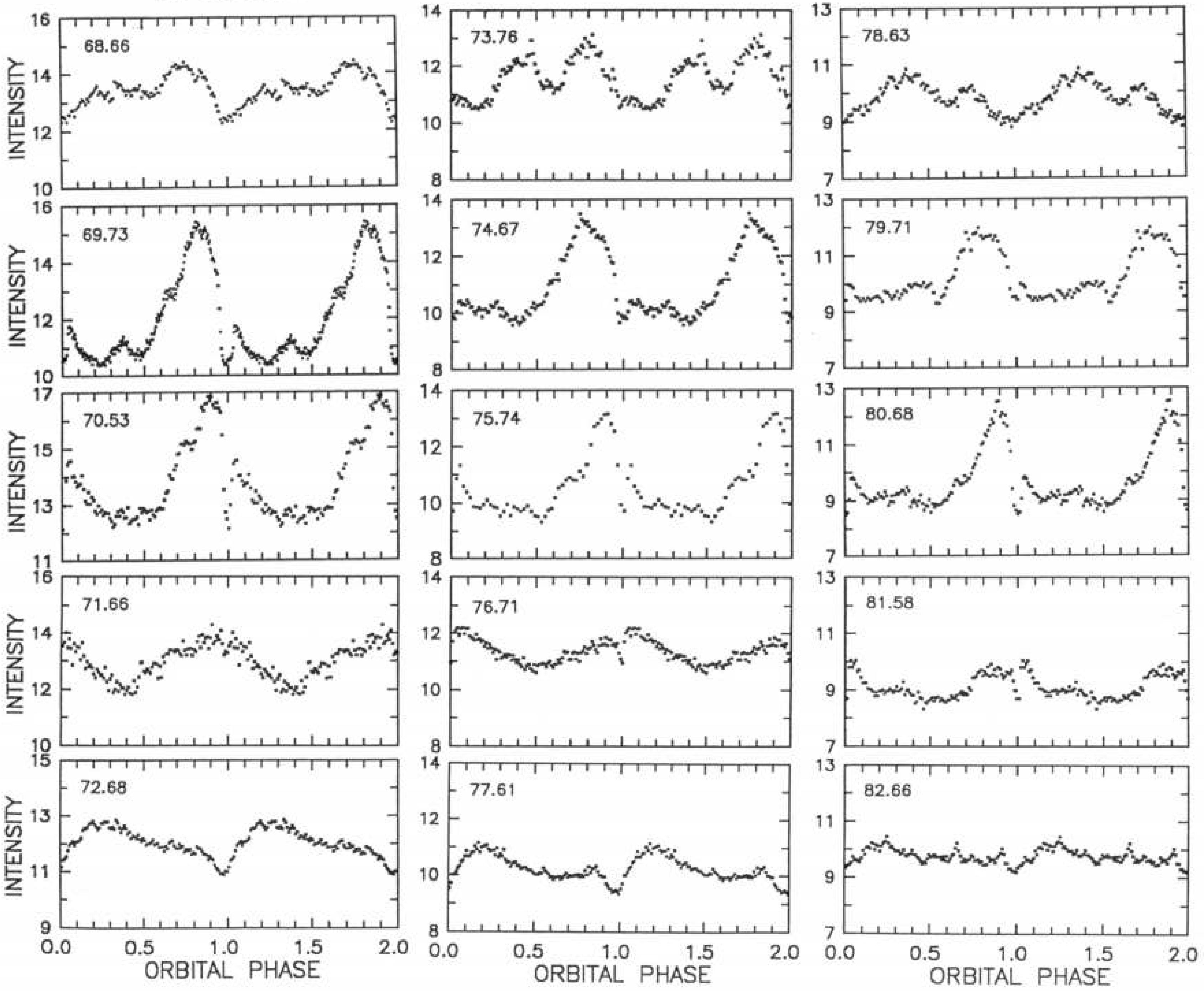

Fig 13

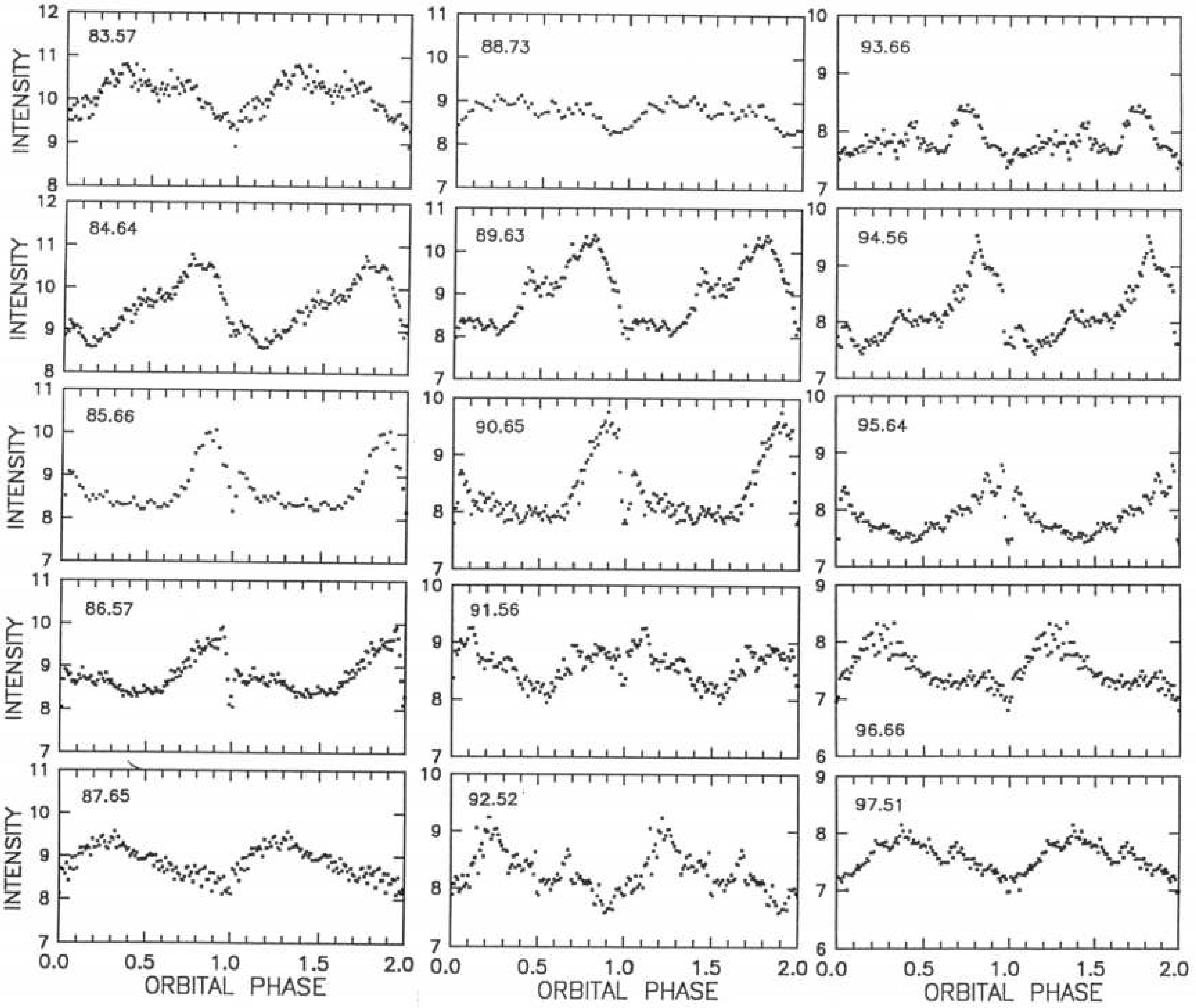

Fig 14

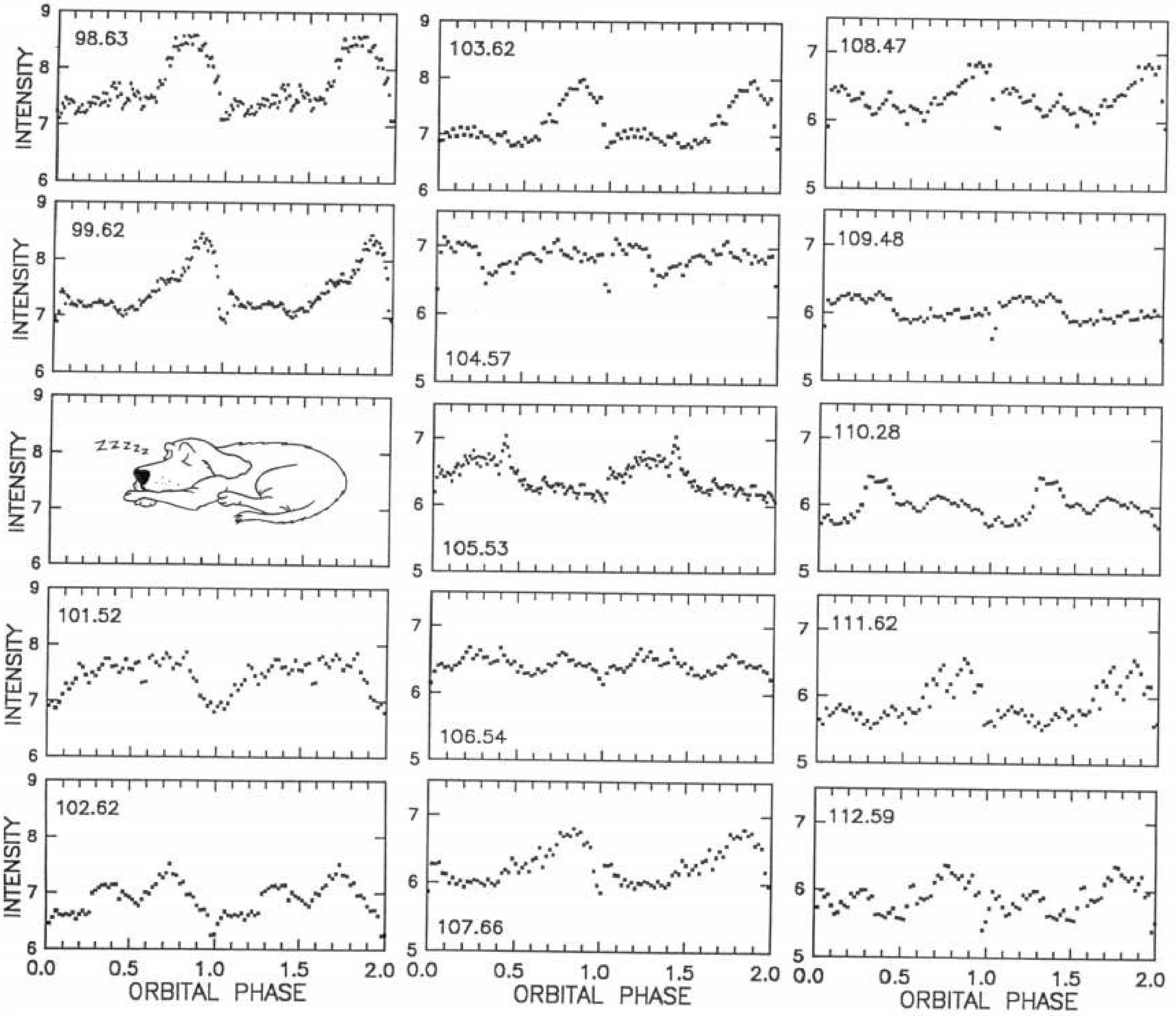

Fig 15

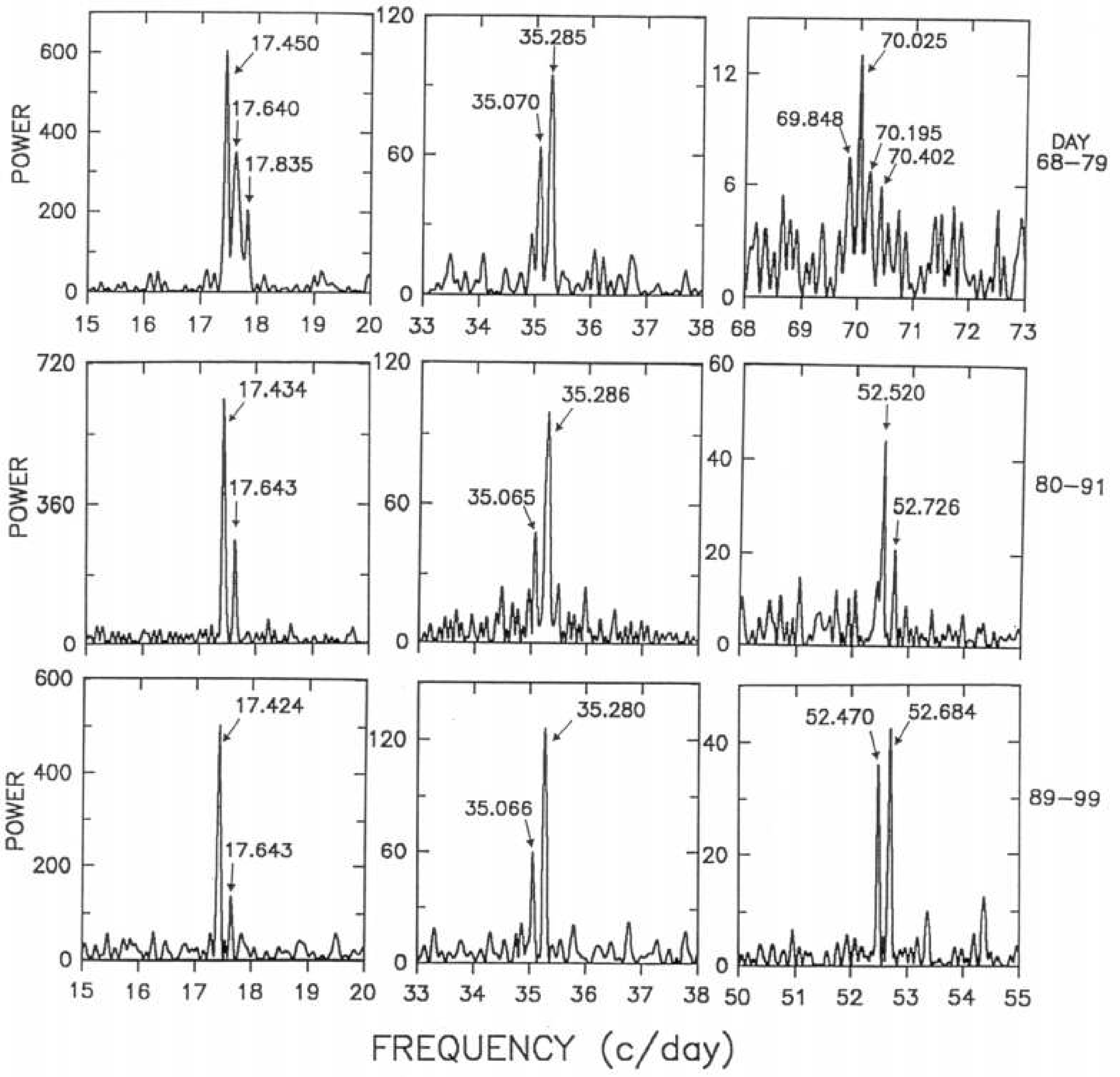

Fig 16

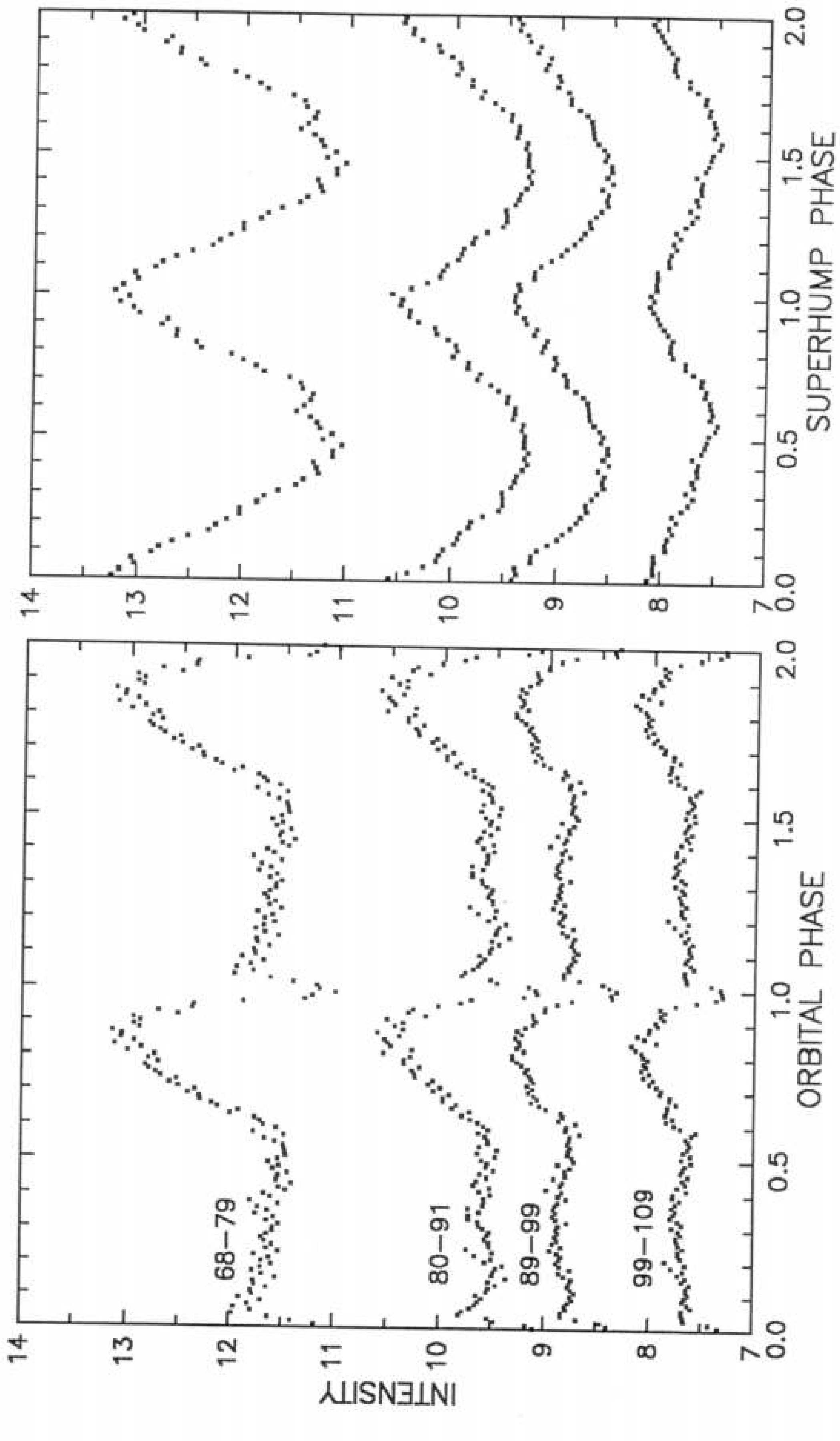

Fig 17

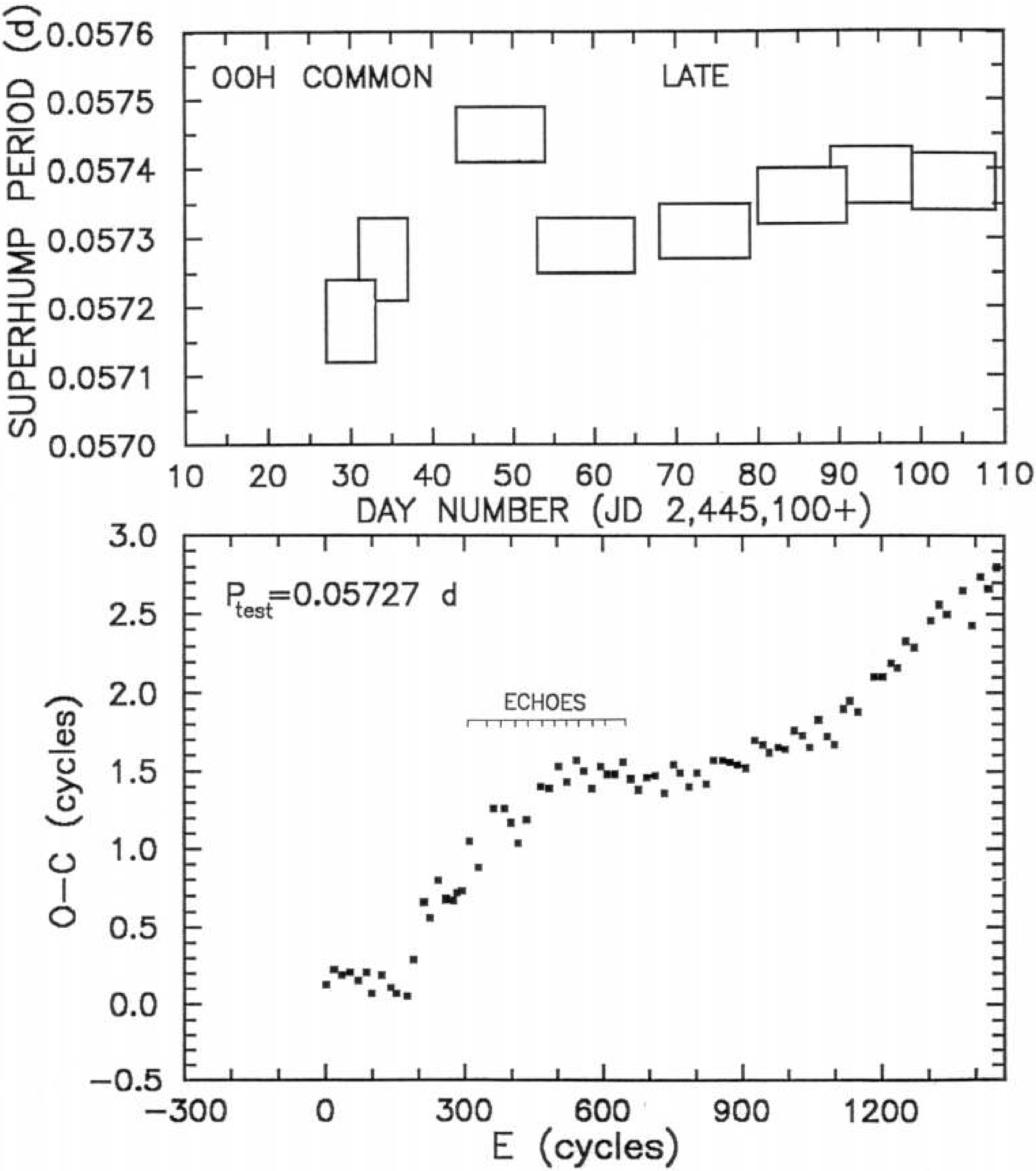

SUPERHUMP PERIOD (d)
0.0576
0.0575
0.0574
0.0573
0.0572
0.0571
0.0570
OOH COMMON
LATE
10 20 30 40 50 60 70 80 90 100 110
DAY NUMBER (JD 2,445,100+)
O−C (cycles)
3.0
2.5
2.0
1.5
1.0
0.5
0.0
−0.5
$P_{test}$=0.05727 d
ECHOES
−300 0 300 600 900 1200
E (cycles)

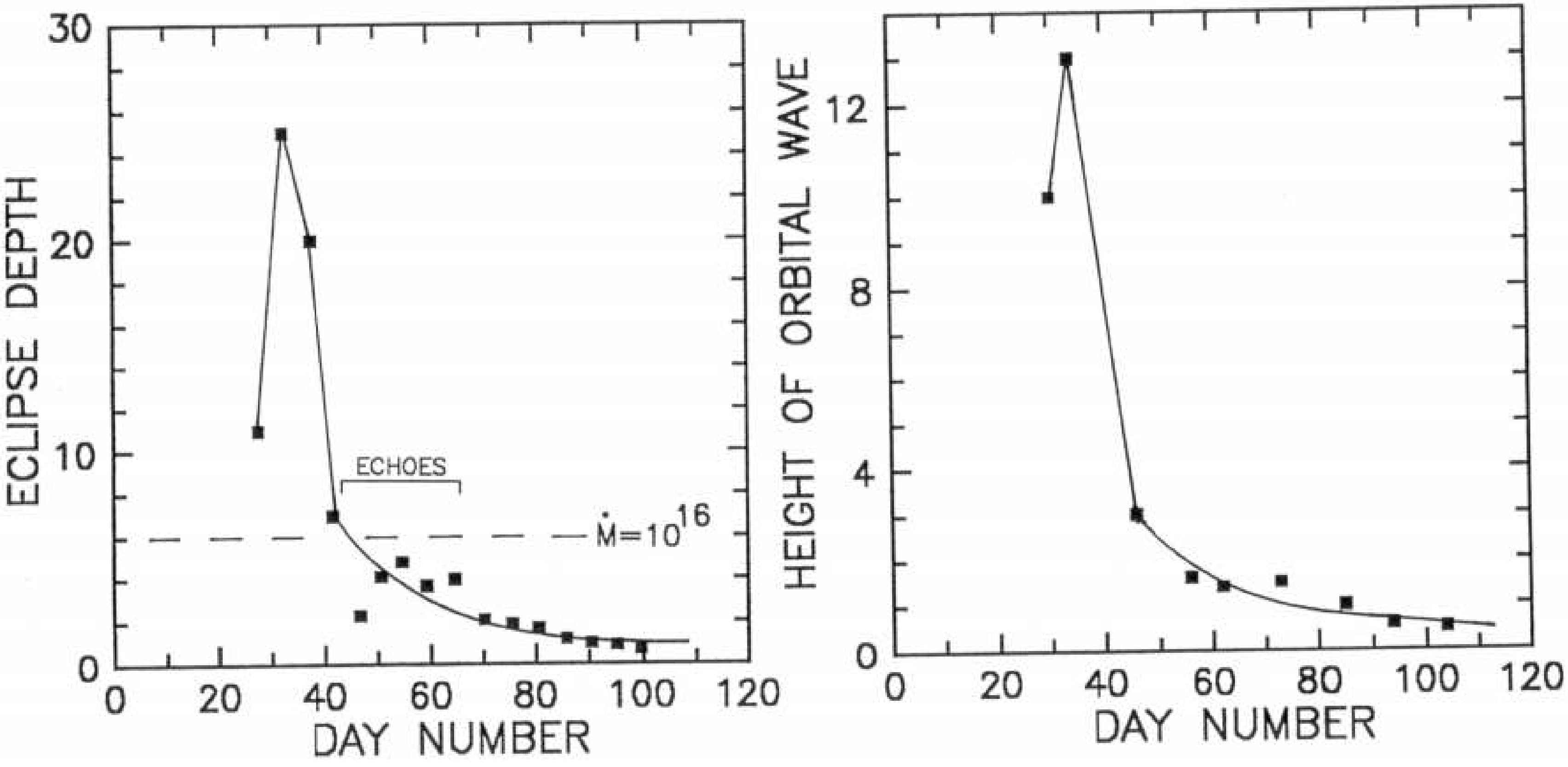

Fig 18

Fig 19

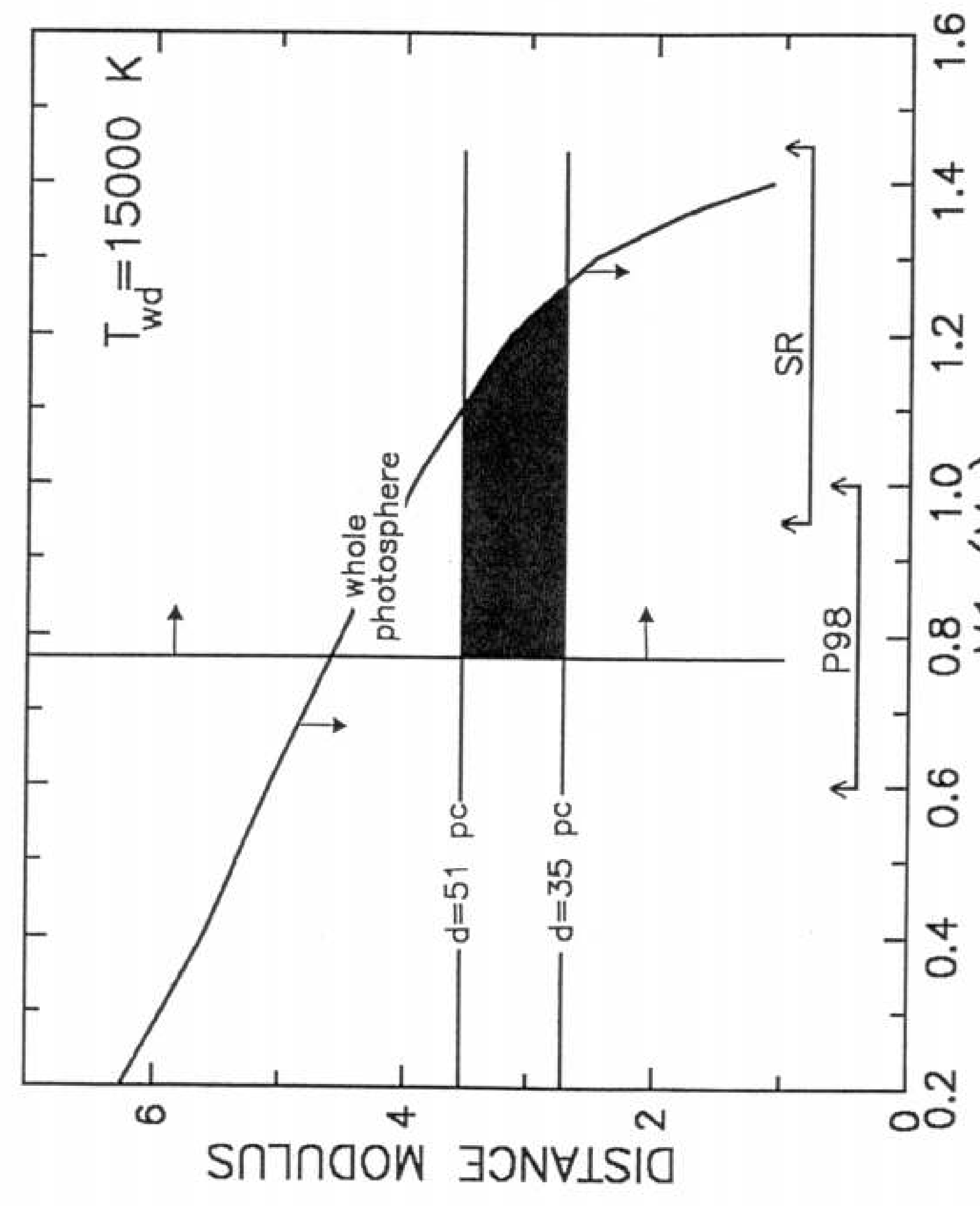

Fig 20

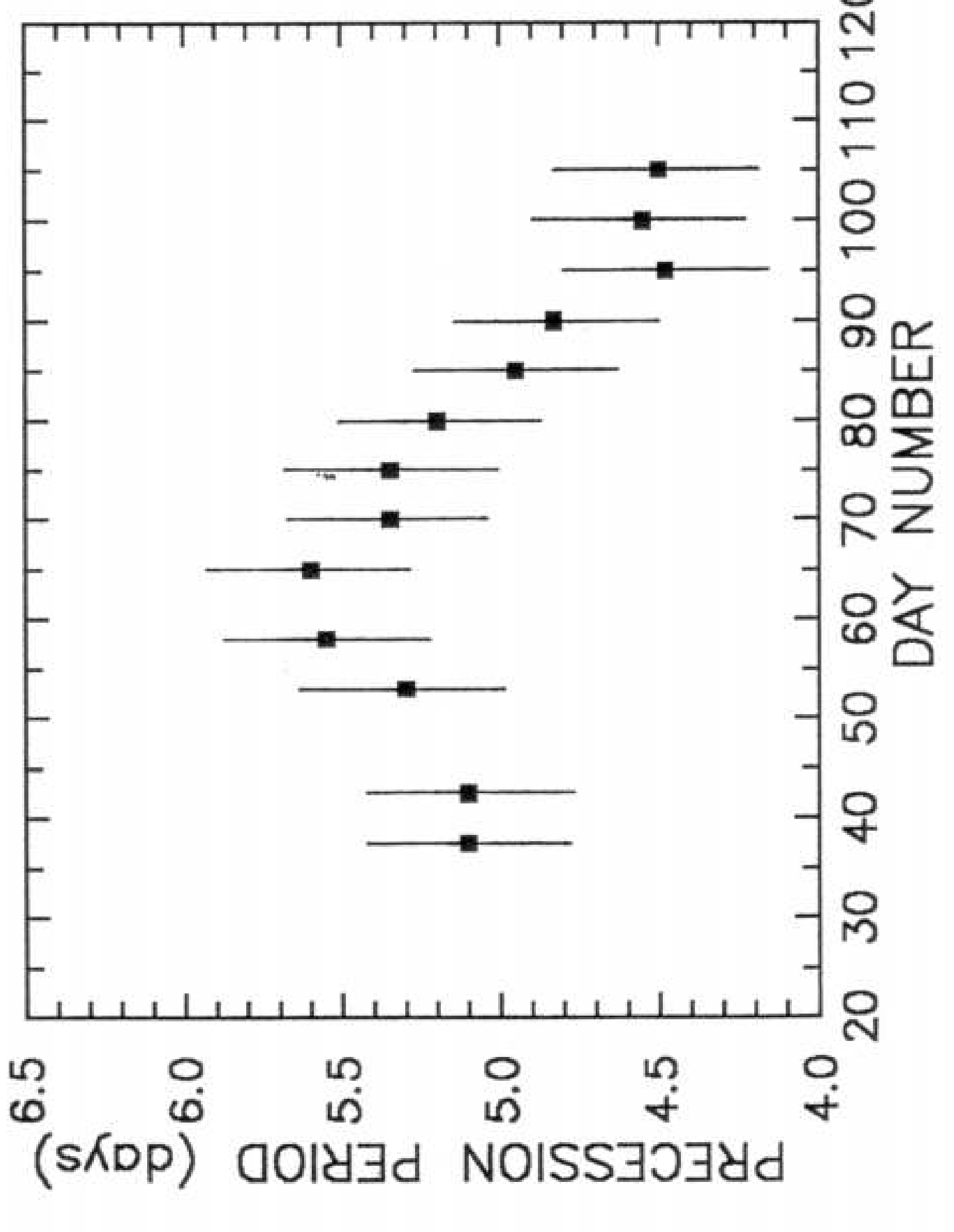

Fig 21

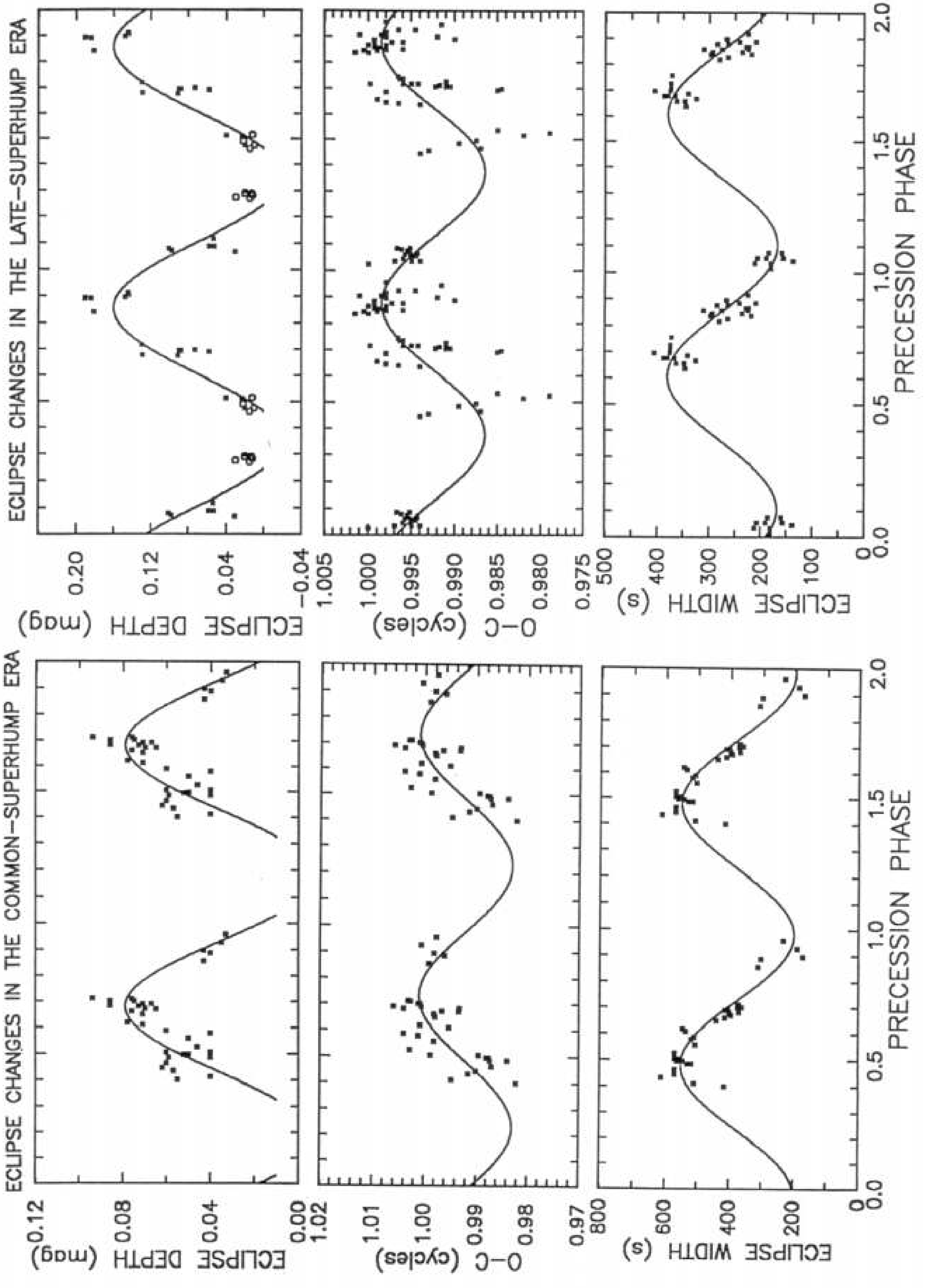

ECLIPSE CHANGES IN THE LATE-SUPERHUMP ERA
ECLIPSE CHANGES IN THE COMMON-SUPERHUMP ERA
ECLIPSE DEPTH (mag)
O-C (cycles)
ECLIPSE WIDTH (s)
PRECESSION PHASE